\documentclass[12pt]{iopart}
\usepackage{iopams}
\usepackage{setstack}
\usepackage{graphicx}

\begin{document}
\title[Bulk viscous matter-dominated model] {Exploring a
matter-dominated model with bulk viscosity to drive the accelerated
expansion of the Universe}

\author{Arturo Avelino and Ulises Nucamendi}

\address{Instituto de F\'{\i}sica y Matem\'aticas\\
Universidad Michoacana de San Nicol\'as de Hidalgo \\
Edificio C-3, Ciudad Universitaria, CP. 58040\\
Morelia, Michoac\'an, M\'exico\\}

\eads{\mailto{avelino@ifm.umich.mx}, \mailto{ulises@ifm.umich.mx}}

\begin{abstract}
We explore the viability of a bulk viscous matter-dominated Universe to explain the present accelerated expansion of the Universe.
The model is composed by a pressureless fluid with bulk viscosity of the form  $\zeta=\zeta_0+\zeta_1 H$ where $\zeta_0$ and $\zeta_1$ are constants and $H$ is the Hubble parameter.
The pressureless fluid characterizes both the baryon and dark matter components.
We study the behavior of the Universe according to this model
analyzing the  scale factor as well as some curvature scalars and the matter density. On the other hand,  we compute the best estimated values of $\zeta_0$ and $\zeta_1$ using the type Ia Supernovae (SNe Ia) probe.
We find that from all the possible scenarios for the Universe, the preferred one by the best estimated values of $(\zeta_0, \zeta_1)$ is that of an expanding Universe beginning with a Big-Bang, followed by a decelerated expansion at early times, and with a smooth transition in recent times to an accelerated expansion epoch that is going to continue forever.
The predicted age of the Universe is a little smaller than the mean value of the observational constraint coming from the oldest globular clusters but it is still inside of the confidence interval of this constraint.
A drawback of the model is the violation of the local second law of
thermodynamics in  redshifts $z \gtrsim 1$.  However, when we assume
$\zeta_1=0$, the simple model $\zeta=\zeta_0$ evaluated at the best
estimated value for $\zeta_0$ satisfies the local second law of
thermodynamics,  the age of the Universe is in perfect agreement
with the constraint of globular clusters, and it also has a
Big-Bang,  followed by a decelerated expansion with the smooth
transition to an accelerated expansion epoch in late times, that is
going to continue forever.
\end{abstract}

 \maketitle

\section{Introduction}
In recent years the type Ia supernovae (SNe Ia) observations have indicated a possibly late time \textit{accelerated expansion} of the Universe \cite{Riess1998}-\cite{Wood-Vasey2007}. This discovery has been supported also by indirect observations of the expansion of the Universe like the cosmic microwave background (CMB) \cite{Bennett2003} and the large scale structure (LSS) \cite{Tegmark2004} observations.
It has been coined the name \textit{dark energy} to refer to the unknown responsible of such acceleration that presumably is the $\sim 70 \% $ of the total content of matter and energy in the Universe \cite{Riess1998}-\cite{Wood-Vasey2007}.

Since the time of this discovery in 1998 \cite{Riess1998}, a large amount of models have been proposed to explain this acceleration. The two most accepted dark energy models are that of a \textit{cosmological constant} (assumed possibly to be the quantum vacuum energy) and a slowly varying rolling scalar field (quintessence models) \cite{Caldwell1998}-\cite{Peebles2003}.

Despite the cosmological constant model is preferred by the observations,
this has several strong problems, among them, the huge
discrepancy between its predicted and observed value (a difference of about
120 orders of magnitude)
\cite{Weinberg1989}-\cite{Padmanabhan} and the problem of that we are living \textit{precisely} in a time when the
matter density in the Universe is of the same order than the dark
energy density (the ``\emph{cosmic coincidence problem}'')
\cite{Zlatev1999}, \cite{Steinhardt1997,Steinhardt1999}.

On the other hand, in the context of \textit{inflation} of the very early Universe, it has been known since long time ago that an imperfect fluid with bulk viscosity in cosmology can produce an acceleration in the expansion of the Universe without the need of a cosmological constant or some inflationary scalar field
\cite{Heller1973}--\cite{Zimdahl1996} (although some authors do not agree with this conclusion \cite{Hiscock1991}). So, extrapolating this theoretical idea used to induce an accelerating Universe without the need of unknown components it is possible to postulate that one candidate to explain the present acceleration can be a bulk viscous pressure \cite{Kremer2003}-\cite{Meng2008}. However, this idea faces some problems too, for example, the need to have a satisfactory mechanism for the origin and composition of the bulk viscosity, nevertheless, there are
already some authors working in this aspect
\cite{Zimdahl2000}-\cite{Mathews2008}.

From a thermodynamical point of view the bulk viscosity in a physical system is due to its deviations from the \textit{local} thermodynamic equilibrium \cite{Wilson} (for a review in the theory of relativistic dissipative fluids see \cite{MaartensHanno}). In a cosmological fluid, the bulk viscosity
may arise when the fluid expands (or contracts) too fast so that the
system does not have enough time to restore its local thermodynamic
equilibrium and then it arises an \emph{effective pressure}
restoring the system to its thermal equilibrium. The
bulk viscosity can be seen as a measurement of this effective
pressure. When the fluid reaches again the thermal equilibrium then
the bulk viscous pressure vanishes \cite{Wilson},
\cite{Okumura}--\cite{Xinzhong}.

So, in an \textit{accelerated} expanding Universe,  it may be
natural to  assume the possibility that the expansion process is
actually a collection of states out of thermal equilibrium in a
small fraction of time giving rise to the existence of a bulk
viscosity.

In the present work we study and test a matter-dominated cosmological model with bulk viscosity as an explanation for the accelerated expansion of the Universe. The model is composed by a pressureless fluid (dust) with bulk viscosity of the form $\zeta = \zeta_0 + \zeta_1 H$ where $\zeta_0$ and $\zeta_1$ are constants to be determined by the observations and $H$ is the Hubble parameter. The pressureless fluid characterizes both the baryon and dark matter components. The term ``$\zeta_0$'' takes into account the simplest parametrization for the bulk viscosity, a constant. And the term ``$\zeta_1 H$'' characterizes the possibility of a bulk viscosity proportional to the expansion ratio of the Universe.

The idea of this model is to drive the present acceleration using the trigger at recent times of
the bulk viscous pressure of the dust fluid instead of any dark energy component.
This fluid represents an unified description of the dark
sector plus the baryon component in a similar way than the Chaplygin
gas model (see for instance \cite{Colistete2007,Jamil2008B} and
references therein).
One of the advantages of the model is that it solves automatically the cosmic coincidence problem because there is not any dark energy component.
The explicit form of the bulk viscosity has to be assumed a priori or obtained from a known or proposed physical
mechanism. In the present work we choose the first option to explore the possibilities, but with the idea to explore the second option in future works.

Different aspects of this model has been also studied in
\cite{Brevik2005AndMore,ArturoUlisesProc2} and references therein.
There is already a large amount of works on viscous fluids in
cosmology with different parametrizations and motivations. For
instance, the particular case of a constant bulk viscosity (the
simplest parametrization) it has been carefully studied in detail
recently \cite{ArturoUlisesPaper1,ArturoUlisesProc1}. Some other
works with the constant parametrization has been also analyzed in
\cite{Heller1973}, \cite{Klimek1974}-\cite{Brevik2002}. Other
parametrizations or more general approaches for the viscous models
have been also proposed and studied at
\cite{Kremer2003}-\cite{Singh2007},
\cite{ChaoJunFeng2009AndMore}-\cite{Hipolito2009}. Another general
approach to the bulk viscous cosmologies called ``fluids with
inhomogeneous equation of state'' have been also proposed and
studied by \cite{Nojiri2005AndMore}.

In section \ref{SectionHubbleParameterWithoutRadiation} we present the main formalism of bulk viscous fluids in General Relativity (GR) and we apply it
to the bulk viscous matter-dominated model studied in this work.
We write the expression of $H$ in terms of the density parameter of the  viscous matter component in function of the redshift. With the expression of $H$ we find the explicit expression for the scale factor that will be used in section \ref{SectionScaleFactor} to analyze the possible scenarios for the Universe according to the values of the \textit{dimensionless}  viscous coefficients $\tilde{\zeta}_0$ and $\tilde{\zeta}_1$.

In sections \ref{SectionCurvature} and \ref{SectionMatterDensity} we
analyze the curvature scalars and the matter density to get a better
understanding of the behavior of the Universe according to the
present model when it is evaluated at the best estimated values of
$\tilde{\zeta}_0$ and $\tilde{\zeta}_1$. In section
\ref{Sectionthermodynamics} we do a brief review of the local second
law of thermodynamics and its possible violation for these models.
Section \ref{SectionSNeTest} presents the SNe Ia probe to be used to
constrain the model and to compute the best estimated values of
$(\tilde{\zeta}_0, \tilde{\zeta}_1)$. Finally, in section
\ref{SectionConclusions} we give our conclusions.

\section{Cosmological model of bulk viscous matter-dominated Universes.}
\label{SectionHubbleParameterWithoutRadiation}

We analyze the cosmological model in a spatially flat Universe framework.
For that, we use the Weinberg formalism \cite{Weinberg} of
imperfect fluids. So, in the present work we consider a bulk
viscous fluid as source of matter in the Einstein fields equations
$G_{\mu \nu} = 8\pi GT_{\mu \nu}$, where $G$ is the Newton
gravitational constant.

The energy-momentum tensor of the bulk viscous matter component is that of
an \textit{imperfect fluid} with a first-order deviation from the thermodynamic equilibrium. It can be expressed as \cite{Wilson,Weinberg}:

\begin{equation}\label{Energy_momentum_tensor}
T^{(\rm m)}_{\mu\nu}=\rho_{\rm m} \,u_\mu u_\nu + (g_{\mu\nu}+u_\mu u_\nu)P^*_{\rm m}\;,
\end{equation}

\noindent where

\begin{equation}\label{pressure}
    P^*_{\rm m} \equiv P_{\rm m} - \zeta \nabla_{\nu}u^{\nu} \;,
\end{equation}

\noindent  $u^\mu$ is the four-velocity vector of an  observer who
measures the energy density $\rho_{\rm m}$ and $P_{\rm m}$  is the
pressure of the fluid of matter, the term $g_{\mu\nu}$ is the metric
tensor, the subscript ``m'' stands for ``matter'' component.
The term $\zeta$ is a \emph{bulk viscous} coefficient  that arises
in a fluid when it is out of the local thermodynamic equilibrium and
that induces a \emph{viscous pressure} equals to $-\zeta
\nabla_{\nu}u^{\nu}$ \cite{Wilson, Okumura}-\cite{Xinzhong}. The
term $P^*_{\rm m}$ is an \emph{effective} pressure composed by the
pressure $P_{\rm m}$ of fluid plus the bulk viscous pressure.

The effective pressure (\ref{pressure}) was  proposed by Eckart \cite{Eckart1940} in 1940 for relativistic dissipative processes in thermodynamics systems out of local equilibrium. Later, Landau \& Lifshitz developed also an equivalent formulation \cite{Landau}.

The Eckart theory has some problems in its formulation, for example, all the equilibrium states in this theory are unstable \cite{Hiscock1985}, other issue is that signals can propagate through the fluids faster than the speed of light \cite{Muller1967,Israel1976}.
To correct the problems of the Eckart theory, Israel-Stewart \cite{IsraelStewart1979A,IsraelStewart1979B}
developed in 1979 a more consistent and general theory that avoids these
issues from which the Eckart theory is the first-order limit when the relaxation time goes to zero. In this limit the Eckart theory is a good approximation to the Israel-Stewart theory.

In spite of the problems of the Eckart theory, but taking advantage
of the equivalence of both theories at this limit, it has been
widely used by several authors because it is simpler to work with
this than with the Israel-Stewart one. In particular, it has been
used to model bulk viscous dark fluids as responsible of the
observed acceleration of the Universe assuming that the
approximation of vanishing relaxation  time is valid for this
purpose (see, for instance
\cite{Kremer2003}-\cite{Brevik2005AndMore},
\cite{Colistete2007,Singh2007,Wilson,ArturoUlisesPaper1,ArturoUlisesProc1}).

It is important to point out that Hiscock \etal  \cite{Hiscock1991}
showed that a flat Friedmann-Robertson-Walker cosmological model
with a bulk viscous Boltzmann gas expands faster when the Eckart
framework is used than with the Israel-Stewart one. He suggests that
the inflationary acceleration due to the bulk viscosity could be
just an effect of using the uncausal Eckart theory and not a real
acceleration. However, posterior studies seem to show that this
conclusion could be wrong because consistent inflationary solutions
have been found using the Israel-Stewart theory \cite{Zimdahl1996}.

The model could have a problem related with the cosmological density perturbations, where a rapid late time decay or irregular perturbations arise, leading to large modifications of the model Lambda Cold Dark Matter ($\Lambda$CDM) predictions for the matter power spectrum, CMB and weak lensing tests, as shown by B. Li et al.  \cite{Baojiu2009}. However, Hipolito et al. \cite{Hipolito2009} have shown that it is possible to have models with bulk viscosity proportional to $\zeta \propto \rho^\nu$ that are compatible with the data from the 2dFGRS and the SDSS surveys. 
Due to the problem with the density perturbations, in the present work we assume that the viscosity appears only till recent times.

It is convenient to mention that for irreversible processes, D. Pav\'on et al. \cite{Pavon1982} have developed a more general formulation than the Israel-Stewart theory, where the temperature could not be associated to the thermal equilibrium. However, in the cosmological scenario, this general formulation has not been totally explored.

Let's start assuming a spatially flat geometry for the
Friedmann-Robertson-Walker (FRW) cosmology as suggested by WMAP
\cite{WMAP}
\begin{equation}\label{FRM_metric}
    ds^2=-dt^2+a^2(t)(dr^2+r^2d\Omega^2)
\end{equation}
where $a(t)$ is the \emph{scale factor}. Moreover, the conservation
equation for the viscous fluid is
\begin{eqnarray}\label{equation1PerfectFluid}
u^{\nu} \nabla_{\nu} \rho_{\rm m} + (\rho_{\rm m} + P^*_{\rm m}) \nabla_{\nu} u^{\nu} &=0
\end{eqnarray}

\noindent We write this conservation equation using the FRW metric as
\begin{eqnarray}
\label{ConservationEquationRadiation}
\dot{\rho}_{\rm m}  + 3H(\rho_{\rm m}  + P^*_{\rm m}) &=0  \label{ConservationEquationViscousComponent}
\end{eqnarray}

\noindent where $H \equiv \dot{a}/a$ is the \emph{Hubble parameter}.
The dot over $\rho_{\rm m}$ and $a$  means time derivative.
For the matter component,  we set  $P_{\rm m}=0$, in concordance
with a \textit{pressureless} fluid. The bulk viscous pressure
$-\zeta \nabla_{\nu}u^{\nu}$ can be written as  $-3\zeta H$. So, the
conservation equation (\ref{ConservationEquationViscousComponent})
for the pressureless viscous fluid becomes

\begin{equation}\label{ConservationEquation-2}
\dot{\rho}_{\rm m}  + 3H(\rho_{\rm m} - 3\zeta H)=0
\end{equation}

\noindent The Friedmann equations for this
model are
\begin{equation}\label{FriedmannFirstEquation}
H^2 = \frac{8\pi G}{3} \rho_{\rm m}
\end{equation}

\begin{equation}\label{FriedmannSecondEquation}
\frac{\ddot{a}}{a}=-\frac{4\pi G}{3} (\rho_{\rm m}  -
9\zeta H)
\end{equation}

\noindent  Using the ansatz $\zeta=\zeta_0+\zeta_1 H$ and the first Friedmann equation (\ref{FriedmannFirstEquation}), we can write the term $(\rho_{\rm m} - 3\zeta H)$ of equation
(\ref{ConservationEquation-2}) as

\begin{equation}\label{ConservationEq-TermAppart}
\rho_{\rm m}-3\zeta H = \left(1-\frac{\tilde{\zeta}_1}{3}
\right) \rho_{\rm m} - \frac{H_0}{(24 \pi G)^{1/2}} \tilde{\zeta}_0 \rho_{\rm m}^{1/2}
\end{equation}

\noindent where we have defined the dimensionless bulk viscous coefficients $\tilde{\zeta}_0$ and $\tilde{\zeta}_1$ as

\begin{equation}\label{DefDimensionlessZetas}
\tilde{\zeta}_0 \equiv \left(\frac{24\pi G}{H_0}\right) \zeta_0, \qquad \tilde{\zeta}_1 \equiv \left(24\pi G\right) \zeta_1
\end{equation}

\noindent and $H_0$ is the Hubble constant. Using the expression (\ref{ConservationEq-TermAppart}) we can rewrite the conservation equation (\ref{ConservationEquation-2}) in terms of the scale factor as

\begin{equation}\label{ODERho-a}
\frac{d \rho_{\rm m}}{da} + \frac{(3-\tilde{\zeta}_1)}{a}\rho_{\rm
m} - \frac{3H_0}{(24 \pi G)^{1/2}} \frac{\tilde{\zeta}_0}{a}\rho_{\rm m}^{1/2} = 0
\end{equation}

\noindent  We divide the ordinary differential equation (ODE) (\ref{ODERho-a}) by the \emph{critical density today}
$\rho^0_{\rm crit} \equiv 3H^2_0 / 8\pi G$ yielding

\begin{equation}\label{ODE_Omega-a}
a\frac{d \hat{\Omega}_{\rm m}}{da} + (3 - \tilde{\zeta}_1) \hat{\Omega}_{\rm m} -
\tilde{\zeta}_0 \hat{\Omega}_{\rm m}^{1/2} =0
\end{equation}

\noindent where we have defined the dimensionless density parameter $\hat{\Omega}_{\rm m}\equiv \rho_{\rm m}/\rho^0_{\rm crit}$. Using the relation between the scale factor $a$ and the redshift $1+z= 1/a$ we can write the ODE (\ref{ODE_Omega-a}) as

\begin{equation}\label{ODE_Omega-z_WithoutR}
(1+z)\frac{d \hat{\Omega}_{\rm m}}{dz} +
(\tilde{\zeta}_1-3)\hat{\Omega}_{\rm m} + \tilde{\zeta}_0
\hat{\Omega}_{\rm m}^{1/2}=0
\end{equation}

\noindent where the function to be calculated in this ODE is
$\hat{\Omega}_{\rm m}(z)$.  This ODE has the exact solution

\begin{equation}\label{ParameterDensity-2}
\hat{\Omega}_{\rm m}(z) = \left[ \left(1 -
\frac{\tilde{\zeta}_0}{3-\tilde{\zeta}_1} \right)
(1+z)^{(3-\tilde{\zeta}_1)/2} +
\frac{\tilde{\zeta}_0}{3-\tilde{\zeta}_1} \right]^2; \quad {\mbox {for}} \: \; \tilde{\zeta}_1 \neq 3
\end{equation}

\noindent where we have used the initial condition $\Omega_{\rm m 0}
\equiv \hat{\Omega}_{\rm m}(z=0) =1$ \footnote{In the present work
we assume $\Omega_{\rm m 0}=1$ like many other authors (see for
instance \cite{Lima2008,Steigman2009} and references therein). To be
consistent with the  dynamical observations measured in clusters of
galaxies that suggest $\Omega_{\rm m 0} \sim 0.26$, it is assumed
that this amount of matter is bounded in structures like galaxies
and cluster of galaxies and that the remaining matter $\Omega_{\rm m
0} \sim 0.74$ is distributed homogeneously in the flat space.} in
concordance with a flat Universe.

On the other hand, we divide the first Friedmann equation
(\ref{FriedmannFirstEquation}) by the critical density today and
then substitute there the expression (\ref{ParameterDensity-2}) to
obtain

\begin{equation}\label{HubbleParameter}
H(z) = H_0 \frac{\left(3 - \tilde{\zeta}_0 - \tilde{\zeta}_1 \right)
(1+z)^{(3-\tilde{\zeta}_1)/2} + \tilde{\zeta}_0}{3-\tilde{\zeta}_1}; \qquad {\mbox {for}} \: \; \tilde{\zeta}_1 \neq 3
\end{equation}

\noindent Or in terms of the scale factor

\begin{equation}\label{HubbleParameterInTermsOfa}
H(a) = H_0 \left[ \left(1 -
\frac{\tilde{\zeta}_0}{3 - \tilde{\zeta}_1} \right)
a^{(\tilde{\zeta}_1-3)/2} +
\frac{\tilde{\zeta}_0}{3-\tilde{\zeta}_1} \right]; \qquad
{\mbox{for}} \; \tilde{\zeta}_1 \neq 3
\end{equation}

        \section{The scale factor.}
        \label{SectionScaleFactor}

We analyze the predicted behavior of the scale factor and the evolution of the Universe according to the present model. So, from the definition of the Hubble parameter $H(a)\equiv \dot{a}/a$ we write the expression (\ref{HubbleParameterInTermsOfa}) as

\begin{equation} \label{ScaleFactorEquation1}
\frac{1}{a} \frac{da}{dt} = H_0 \left[ \left(1 -
\frac{\tilde{\zeta}_0}{3-\tilde{\zeta}_1} \right)
a^{(\tilde{\zeta}_1-3)/2} +
\frac{\tilde{\zeta}_0}{3-\tilde{\zeta}_1} \right]
\end{equation}

\noindent We define $\gamma \equiv \tilde{\zeta}_0/(3-\tilde{\zeta}_1)$ and $\beta \equiv \tilde{\zeta}_1 - 3$, and integrate the equation
(\ref{ScaleFactorEquation1}) as follows

\begin{equation}
\int^a_1 \frac{da'}{a' \left[(1-\gamma)a'^{\beta/2} + \gamma
\right]} = \int^t_{t_0} H_0 dt' = H_0 (t-t_0)
\end{equation}

\noindent where $t$ is the \emph{cosmic time}. Solving this integral we obtain

\begin{equation}
- \frac{2}{\gamma \beta} \ln\left[1-\gamma +
\frac{\gamma}{a^{\beta/2}} \right] = H_0 (t-t_0)
\end{equation}

\noindent Solving for $a(t)$ and substituting the values of $\gamma$ and $\beta$ we arrive to

\begin{equation}\label{ScaleFactorExpression}
\fl a(t) = \left[\frac{\tilde{\zeta}_0+\tilde{\zeta}_1-3 + (3-\tilde{\zeta}_1) \exp
\left( \frac{1}{2} \tilde{\zeta}_0 H_0 (t-t_0)
\right)}{\tilde{\zeta}_0} \right]^{2/(3-\tilde{\zeta}_1)}; \quad {\mbox {for}} \: \tilde{\zeta}_0 \neq 0, \; \tilde{\zeta}_1 \neq 3
\end{equation}

Also, we compute the first and second derivatives of expression (\ref{ScaleFactorExpression}) with respect to $x\equiv H_0 (t-t_0)$ to study the accelerated or decelerated epochs of the Universe and the transitions between them

\begin{equation}\label{ScaleFactor1DeriveZ1Neq3}
\frac{da}{dx} = \e^{\tilde{\zeta}_0 x /2} \left[ \frac{\tilde{\zeta}_0 + \tilde{\zeta}_1 -3 + (3-\tilde{\zeta}_1) \e^{\tilde{\zeta}_0 x /2} }{\tilde{\zeta}_0} \right]^{(\tilde{\zeta}_1 - 1)/(3 - \tilde{\zeta}_1)},
\end{equation}

\begin{equation}\label{ScaleFactor2DeriveZ1Neq3}
\fl \frac{d^2 a }{d x^2} = \frac{1}{2} \e^{\tilde{\zeta}_0 x /2} \left(\tilde{\zeta}_0 + \tilde{\zeta}_1 -3 + 2\e^{\tilde{\zeta}_0 x /2}  \right)\left[\frac{\tilde{\zeta}_0 + \tilde{\zeta}_1 -3 + (3-\tilde{\zeta}_1) \e^{\tilde{\zeta}_0 x /2} }{\tilde{\zeta}_0} \right]^{2(\tilde{\zeta}_1-2)/(3-\tilde{\zeta}_1)}
\end{equation}

We have analyzed the behavior of the scale factor
in all the possible combinations of values  for
$(\tilde{\zeta}_0,\tilde{\zeta}_1)$. We have found that from all the
possibilities, the cases where there is a Big-Bang as the origin of
the Universe, followed by a \textit{decelerated} expansion epoch at
early times and with a \textit{transition} to an
\textit{accelerated} epoch in recent times, corresponds to the cases
with values $(\tilde{\zeta}_0 > 0, \; \tilde{\zeta}_1 < 0)$ and $(0<
\tilde{\zeta}_0 <3, \; \tilde{\zeta}_1 = 0)$. We have realized also that
the best estimated values for $(\tilde{\zeta}_0,\tilde{\zeta}_1)$,
or $\tilde{\zeta}_0$ alone (when we set $\tilde{\zeta}_1=0$)
computed using the latest SNe Ia data set, lie precisely in these
two ranges of values respectively [see table
\ref{tableSummary_Z01}]. We present below the details of the analysis of these two cases. 

It is important to highlight that  in general for any other case
different of the two cases $(\tilde{\zeta}_0>0, \tilde{\zeta}_1<0)$
(with $\tilde{\zeta}_0+\tilde{\zeta}_1<3$), and
$(0<\tilde{\zeta}_0<3, \tilde{\zeta}_1=0)$, the Universe does not
have the same behavior, i.e., for some other cases there is not a
Big-Bang or a late time accelerated epoch with a early time
deceleration period, etc.

First, we describe the case $(\tilde{\zeta}_0 > 0, \; \tilde{\zeta}_1 < 0)$.
There are two kinds of behaviors depending on the inequality:

\begin{equation}
\mbox{\textbf{(a)}} \; \tilde{\zeta}_0 + \tilde{\zeta}_1 < 3, \qquad \qquad
\mbox{\textbf{(b)}} \; \tilde{\zeta}_0 + \tilde{\zeta}_1 > 3
\end{equation}

\begin{figure}
\begin{center}
\includegraphics[width=10cm]{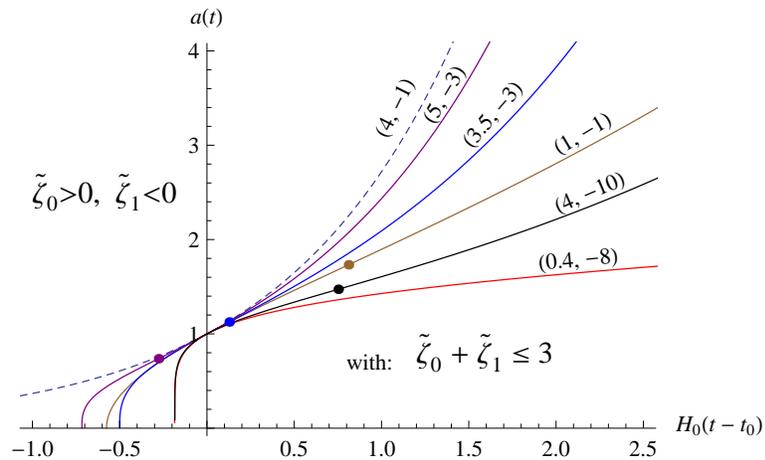}
\caption{Plot of the scale factor with respect to $H_0(t-t_0)$ for
different values of $(\tilde{\zeta}_0, \tilde{\zeta}_1)$ when
$\tilde{\zeta}_0 > 0$ and $\tilde{\zeta}_1 < 0$ but with the
condition  $\tilde{\zeta}_0 + \tilde{\zeta}_1 \leq 3$ [see
expression (\ref{ScaleFactorExpression})]. The model predicts an
\textit{eternally expanding} Universe that begins with a Big-Bang
followed by a decreasing \textit{decelerated} expansion until a time
when the deceleration vanishes and then there is a
\textit{transition} to an \textit{accelerated} expansion epoch that
is going to continue \textit{forever}, so that $a \rightarrow
\infty$ when $t \rightarrow \infty$.  See section
\ref{SectionScaleFactor} for details. The points indicate the time
when the transition between the \textit{decelerated expansion} epoch
to  an \textit{accelerated} one happens [see expression
(\ref{transitionDecAceZ0Gr0Z1Ls0})]. The dashed line corresponds to
a \textit{de~Sitter Universe} that is obtained when $\tilde{\zeta}_0
+ \tilde{\zeta}_1=3$.} \label{PlotScaleFactorZ0Gr0Z1Ls0}
\end{center}
\end{figure}

\begin{figure}
\begin{center}
\includegraphics[width=16cm]{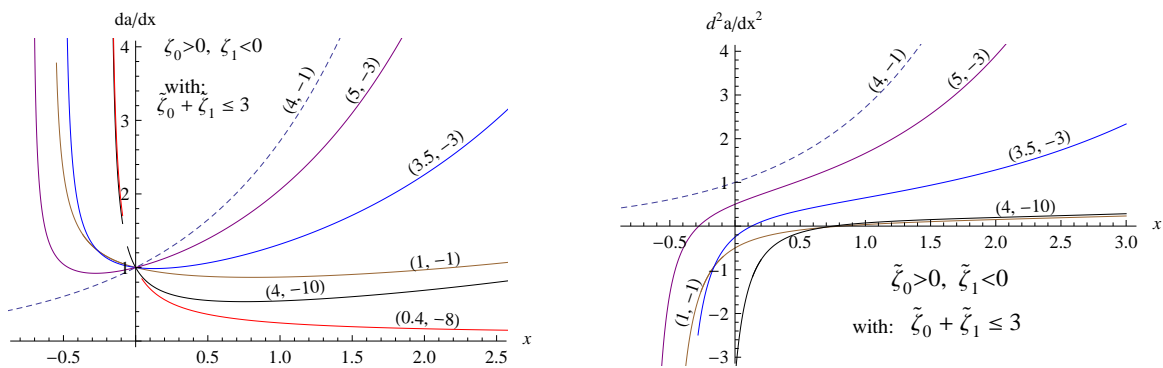}
\caption{Plot of the first and second derivative of the scale factor with respect to $x\equiv
H_0(t-t_0)$ for different values of $(\tilde{\zeta}_0, \tilde{\zeta}_1)$ when $\tilde{\zeta}_0 > 0$ and $\tilde{\zeta}_1 < 0$ but with the condition  $\tilde{\zeta}_0 + \tilde{\zeta}_1 \leq 3$  [see equations
(\ref{ScaleFactor1DeriveZ1Neq3}) and (\ref{ScaleFactor2DeriveZ1Neq3})].
These plots show  the decelerated and accelerated expanding epochs as well as the transition between them (see section \ref{SectionScaleFactor}). For the left figure, the \textit{minimum} value of each plot corresponds to the cosmic time when the transition happens, that for the right figure corresponds when a plot intercepts the $x$-axis.} \label{PlotScaleFactor12DeriveZ0Gr0Z1Ls0}
\end{center}
\end{figure}

\begin{figure}
\begin{center}
\includegraphics[width=10cm]{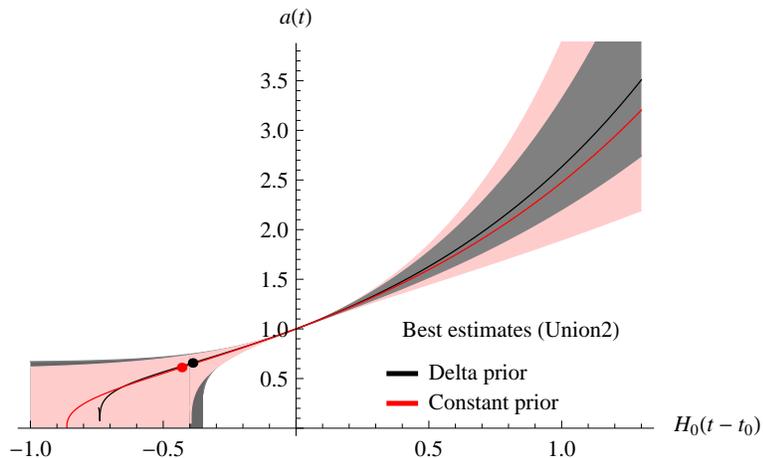}
\caption{Plot of the scale factor $a(t)$ with respect to
$H_0(t-t_0)$ [see expression (\ref{ScaleFactorExpression})] when it is evaluated at the best estimated values of $\tilde{\zeta}_0$ and $\tilde{\zeta}_1$ (see table \ref{tableSummary_Z01}). The black line corresponds to the values $(\tilde{\zeta}_{\rm 0 }=8.58 \pm 1.2, \;
\tilde{\zeta}_{\rm 1}=-5.96 \pm 1.1)$ coming from the use of a
\emph{Dirac Delta prior} distribution to marginalize over the Hubble constant $H_0$.  The red line corresponds to the values
$(\tilde{\zeta}_{\rm 0}=4.38 \pm 1.5, \; \tilde{\zeta}_{\rm 1
}=-2.16 \pm 1.4)$ coming from the use of a \emph{Constant prior} distribution to marginalize over $H_0$ (see table
\ref{tableSummary_Z01}).
The plot shows how the model, evaluated at the best estimates, predicts a Big-Bang as origin of the Universe, followed by a \textit{decelerated} expanding epoch at early times, with a \textit{transition} to an accelerated epoch at late times that will continue forever.
The points indicate the cosmic time when the transition
between the decelerated epoch to an accelerated one happens [see expression (\ref{transitionDecAceZ0Gr0Z1Ls0}) and table \ref{tableAgeTransition}]. The
grey and pink bands correspond to evaluate the expression
(\ref{ScaleFactorExpression}) at the errors in the estimations for
bulk viscous coefficients, i.e., for the Dirac delta case (pink
band) the expression (\ref{ScaleFactorExpression}) was evaluated at
$(\tilde{\zeta}_0=8.58+1.2, \tilde{\zeta}_1=-5.96 + 1.1)$ to plot
the upper limit of the band and $(\tilde{\zeta}_0=8.58-1.2,
\tilde{\zeta}_1=-5.96 - 1.1)$  for the lower one. In the same way,
for the constant prior (grey band) $(\tilde{\zeta}_{\rm 0}=4.38 \pm
1.5,\tilde{\zeta}_{\rm 1}=-2.16 \pm 1.4)$ for the upper and lower limits
respectively.} \label{PlotScaleFactorBestEstimate}
\end{center}
\end{figure}

\begin{figure}
\begin{center}
\includegraphics[width=15cm]{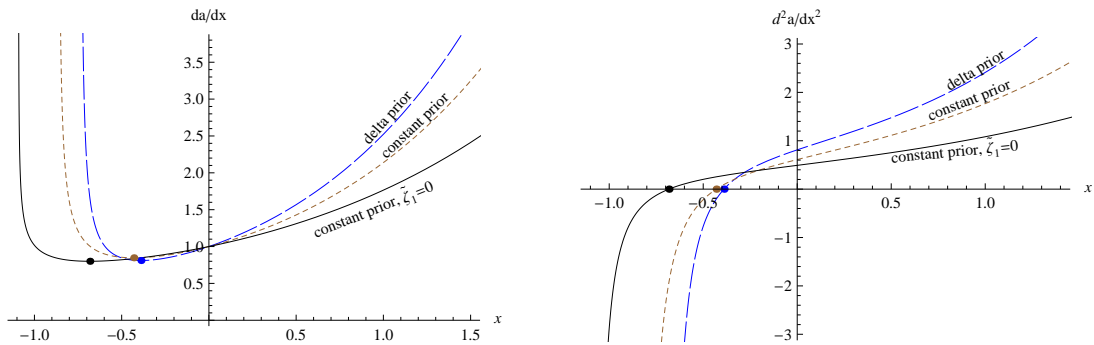}
\caption{Plot of the first and second derivative of the scale factor $a(t)$ with respect to $x \equiv H_0(t-t_0)$ [see expressions  (\ref{ScaleFactor1DeriveZ1Neq3}) and (\ref{ScaleFactor2DeriveZ1Neq3})] when they are evaluated at the best estimated values of $\tilde{\zeta}_0$ and $\tilde{\zeta}_1$ (see table \ref{tableSummary_Z01}).
The points indicate the time when the transition between the \textit{decelerated expanding} epoch to an \textit{accelerated} one happens [see expression (\ref{transitionDecAceZ0Gr0Z1Ls0}) and table \ref{tableAgeTransition}].
For the left figure, the \textit{minimum} value of each plot corresponds to the cosmic time when the transition happens, that for the right figure corresponds when a plot intercepts the $x$-axis.
The legends ``\textit{delta prio}r'', ``\textit{constant prior}'' and ``\textit{constant prior},  $\tilde{\zeta}_1=0$'' label the different marginalizations used to compute the best estimated values (see table \ref{tableSummary_Z01}).}
\label{PlotScaleFactorBest12Derive}
\end{center}
\end{figure}

\begin{figure}
\begin{center}
\includegraphics[width=10cm]{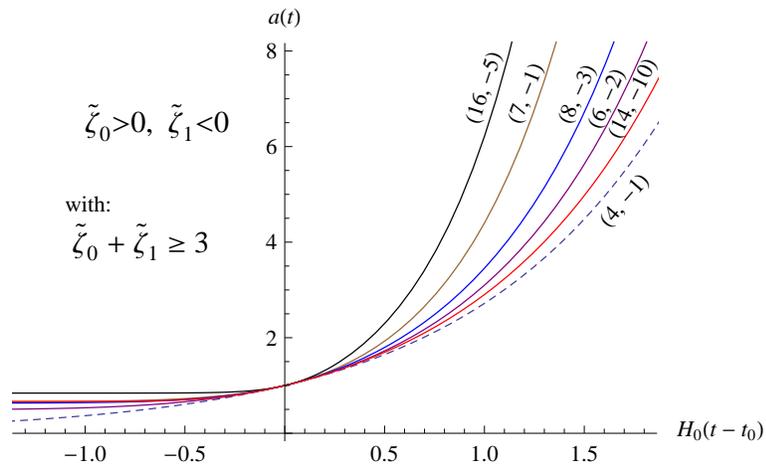}
\caption{Plot of the scale factor  with respect to $H_0(t-t_0)$ for
different values of $(\tilde{\zeta}_0, \tilde{\zeta}_1)$ when
$\tilde{\zeta}_0 > 0$ and $\tilde{\zeta}_1 < 0$ but with the
condition of $\tilde{\zeta}_0 + \tilde{\zeta}_1 \geq 3$  [see
equation (\ref{ScaleFactorExpression})]. The model predicts an
eternally expanding Universe and this expansion is always
accelerated. \textit{There is not} a Big-Bang as the origin of the
Universe, instead of that, the Universe has a \textit{minimum} value
of the scale factor given by $a_{\rm min}  = [(\tilde{\zeta}_0 +
\tilde{\zeta}_1 - 3)/\tilde{\zeta}_0 ]^{2/(3 - \tilde{\zeta}_1)}$
when $t \rightarrow -\infty$.  From this minimum value $a_{\rm min}$
the scale factor increases along the time until $a \rightarrow
\infty$ when $t \rightarrow \infty$. See section
\ref{SectionScaleFactor} for details.}
\label{PlotScaleFactorZ0Gr0Z1Ls0Gr3}
\end{center}
\end{figure}

Figures
\ref{PlotScaleFactorZ0Gr0Z1Ls0}--\ref{PlotScaleFactorZ0Gr0Z1Ls0Gr3}
show the behavior of the scale factor for these subcases  using
different values for $(\tilde{\zeta}_0, \tilde{\zeta}_1)$.

For the  subcase \textbf{(a)} the model predicts an
\textit{eternally expanding} Universe beginning with a
Big-Bang\footnote{In the present work we assume as a
\textit{Big-Bang} the state of the Universe where the scale factor
is zero in some cosmic time $t_{\rm BB}$ in the past of the
Universe, i.e., $a(t_{\rm BB}) = 0$, where in addition, the curvature scalar $R$
and the matter density $\rho$ are singulars, i.e., $R(a=0) =
\rho(a=0) =\infty$ (see sections \ref{SectionCurvature} and
\ref{SectionMatterDensity} for details). The subscript ``BB'' stands
for ``Big-Bang''.} in the past followed by a \textit{decelerated}
expansion where the deceleration is \textit{decreasing} until a time
when it vanishes and then there is a smooth \textit{transition} to
an \textit{accelerated} expansion epoch that is going to continue
\textit{forever}, so that $a \rightarrow \infty$ when $t \rightarrow
\infty$ (see figures
\ref{PlotScaleFactorZ0Gr0Z1Ls0}-\ref{PlotScaleFactorBest12Derive}).

The cosmic time of \textit{transition} ``$t_{\rm trans}$'' between the decelerated to the accelerated expansion epochs can be calculated equating to zero the expression (\ref{ScaleFactor2DeriveZ1Neq3}) yielding

\begin{equation}\label{transitionDecAceZ0Gr0Z1Ls0}
t_{\rm trans} = t_0 + \frac{2}{H_0 \tilde{\zeta}_0} \ln \left( \frac{3  - \tilde{\zeta}_0- \tilde{\zeta}_1}{2} \right)
\end{equation}

To find the expression of the transition between the decelerated to the accelerated expansion epochs in terms of the scale factor we derive with respect to $a$ the expression (\ref{HubbleParameterInTermsOfa}) and using the fact of that $H \equiv \dot{a}/a$ we obtain

\begin{equation}\label{ProcATransition}
\frac{d\dot{a}}{da} = H_0 \left[\frac{\tilde{\zeta}_1-1}{2} \left(1-\frac{\tilde{\zeta}_0}{3-\tilde{\zeta}_1} \right)a^{(\tilde{\zeta}_1-3)/2} + \frac{\tilde{\zeta}_0}{3-\tilde{\zeta}_1} \right]
\end{equation}

\noindent We equal to zero the equation (\ref{ProcATransition}) yielding

\begin{equation}\label{DefScaleFactorTransition}
a_t = \left[\frac{2\tilde{\zeta}_0}{(\tilde{\zeta}_1-1)(\tilde{\zeta}_0+\tilde{\zeta}_1-3)} \right]^{2/(\tilde{\zeta}_1-3)}
\end{equation}

\noindent The transition happens in the \textit{past} of the Universe if $1<\tilde{\zeta}_0 + \tilde{\zeta}_1 <3$, in the \textit{future} if
$\tilde{\zeta}_0 + \tilde{\zeta}_1 <1$  and \textit{today} if $\tilde{\zeta}_0 + \tilde{\zeta}_1 =1$.

On the other hand, using $a(t_{\rm BB})=0$, we find the cosmic time $t_{\rm BB}$ when the Big-Bang happens

\begin{equation}\label{BigBangTimeZ0Ls0Z1Ls0}
t_{\rm BB}=t_0 + \frac{2}{H_0 \tilde{\zeta}_0} \ln \left(1 - \frac{\tilde{\zeta}_0}{3 - \tilde{\zeta}_1} \right)
\end{equation}

We define the \textit{Age of the Universe} as the elapsed time between the time $t_{\rm BB}$ until the present time $t_0$. So, using the expression (\ref{BigBangTimeZ0Ls0Z1Ls0}), the age of the Universe  is given by

\begin{equation}\label{AgeUniverseZ0Ls0Z1Ls0}
{\mbox{Age}} \equiv |t_0 - t_{\rm BB}| =  \left| \frac{2}{H_0 \tilde{\zeta}_0} \ln \left(1 - \frac{\tilde{\zeta}_0}{3 - \tilde{\zeta}_1} \right) \right|
\end{equation}

The best estimated values for the bulk viscous dimensionless coefficients $(\tilde{\zeta}_0, \tilde{\zeta}_1)$ computed using the type Ia SNe data correspond to this subcase (see table \ref{tableSummary_Z01} and section \ref{SectionSNeTest} for details). The figures  \ref{PlotScaleFactorBestEstimate} and \ref{PlotScaleFactorBest12Derive} show how the model, evaluated at the best estimates, predicts a Big-Bang as origin of the Universe, followed by a \textit{decelerated} expanding epoch at early times, with a \textit{transition} to an accelerated epoch at late times that will continue forever.

Table \ref{tableAgeTransition} shows the values for the age of the Universe, evaluated at the best estimates, as well as the value of the redshift of transition between deceleration-acceleration epochs. From table~\ref{tableAgeTransition} we can see that for the case of both $(\tilde{\zeta}_0, \tilde{\zeta}_1)$ as free parameters (i.e., when $\tilde{\zeta}_1 \neq 0$), and for both marginalizations over the Hubble constant $H_0$, the estimated values for the age of the Universe (11.72 and 10.03 Gyr) are smaller  than the mean value of $12.9$ Gigayears coming from the oldest globular cluster observations but still inside of the confidence interval of these: $12.9 \pm 2.9$ Gigayears (see section \ref{SectionSNeTest} and table~\ref{tableAgeTransition} for details).

For the subcase  \textbf{(b)} the model predicts an eternally
expanding Universe and this expansion is always accelerated along
the whole history of the Universe  (see figure
\ref{PlotScaleFactorZ0Gr0Z1Ls0Gr3}). In this subcase \textit{there is not} a
Big-Bang as the origin of the Universe, instead of that, the
Universe has a \textit{minimum} value of the scale factor given by

\begin{equation}\label{ScaleFactorbUZ0Gr0Z1Ls0Gr3}
a_{\rm min} \equiv \lim_{t \rightarrow -\infty} a(t) = \left(\frac{\tilde{\zeta}_0
+ \tilde{\zeta}_1 - 3}{\tilde{\zeta}_0}
\right)^{2/(3 - \tilde{\zeta}_1)}
\end{equation}

\noindent where  the subscript ``min'' stands for ``minimum''. From
this minimum value $a_{\rm min}$ the scale factor increases along
the time until $a \rightarrow \infty$ when $t \rightarrow \infty$.
For this subcase the age of the Universe is not defined [see
equation (\ref{AgeUniverseZ0Ls0Z1Ls0})].

\begin{table}
  \centering
\begin{tabular}{| l | c | c c |}
\multicolumn{4}{c}{\textbf{Viscous model $\zeta=\zeta_0 + \zeta_1 H$}}\\

\hline Prior over $H_0$ & Age Universe & $a_t$ & $z_t$ \\
\hline

Constant & 11.72 Gyr& 0.61 & 0.63 \\
Dirac delta & 10.03 Gyr& 0.65 & 0.52 \\
Constant ($\tilde{\zeta}_1=0$) & 14.82 Gyr& 0.40 & 1.47 \\
\hline

\end{tabular}
\caption{Summary of the computed central values of the age of the Universe, the scale factor of transition between deceleration-acceleration epochs ``$a_t$'' and its corresponding value in redshift ``$z_t$'', coming from the best estimates of $\tilde{\zeta}_0$ and $\tilde{\zeta}_1$ [see table \ref{tableSummary_Z01}]. The first column shows the prior distribution assumed for $H_0$ to marginalize. The second column shows the estimated central values for the age of the Universe [see expression (\ref{AgeUniverseZ0Ls0Z1Ls0})] given in units of Gigayears (Gyr). These magnitudes were obtained assuming a value of $H_0= 72 \; ({{\rm km}}/{{\rm s}}){{\rm Mpc}}^{-1}$ in the expression (\ref{AgeUniverseZ0Ls0Z1Ls0}) as suggested by the observations of
the Hubble Space Telescope (HST) \cite{Freedman2001}, assuming a value of a year of 31556925.2 seconds (a \textit{tropical} year) and a megaparsec $=3.0856776 \times 10^{19}$ km. The third column shows the estimated
value for scale factor $a_t$  when the transition between the
\emph{decelerated} expanding epoch to the \emph{accelerated}
expanding one happens [see expression (\ref{DefScaleFactorTransition})].
The fourth column shows also the
value of this transition expressed in terms of the redshift and
labeled as $z_t$ (the relation is given by $1+z= 1/a$).}
\label{tableAgeTransition}
\end{table}

\begin{figure}
\begin{center}
\includegraphics[width=10cm]{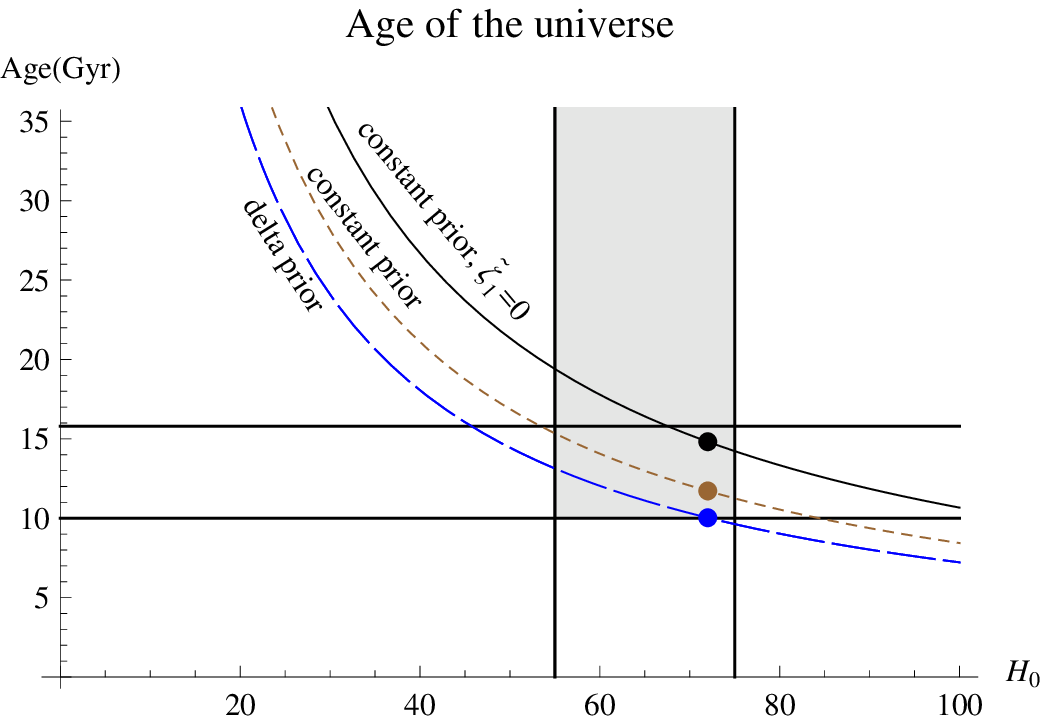}
\caption{Plot of the age of the Universe in terms of the Hubble
constant $H_0$ when it is evaluated at the best estimated values for $(\tilde{\zeta}_0, \tilde{\zeta}_1)$ [see expression (\ref{AgeUniverseZ0Ls0Z1Ls0}) and tables \ref{tableAgeTransition} and \ref{tableSummary_Z01}].
$H_0$ is given in units of $({{\rm km}}/{{\rm s}}){{\rm Mpc}}^{-1}$ and the age in units of Gigayears (Gyr).
It was assumed a value of a year of 31556925.2 seconds (a \textit{tropical} year) and a megaparsec $=3.0856776 \times 10^{19}$ km.
The points indicate the computed value for the age of the Universe when it is assumed a value of $H_0=72\,({{\rm km}}/{{\rm
s}}) {{\rm Mpc}}^{-1}$, as suggested by the observations of
the Hubble Space Telescope (HST) \cite{Freedman2001}.
The vertical lines correspond to the interval $H_0=[55,75]\,({{\rm km}}/{{\rm s}}) {{\rm Mpc}}^{-1}$, it is the permitted region according to values of $H_0$ consistent with the distance moduli used to derive ages for Galactic globular clusters from the \emph{Hipparcos} parallaxes (see ref. \cite{CarretaEtal2000}). The horizontal lines correspond to the constraint for the age of the Universe from the oldest globular clusters (Age$=12.9 \pm 2.9$ Gyr \cite{CarretaEtal2000}). So, the shaded area is the consistent region for the age of the Universe. Table \ref{tableAgeTransition} summarizes the estimated values for the age of the Universe.}
\label{PlotAgeUniverseSNeU10}
\end{center}
\end{figure}

\begin{figure}
\begin{center}
\includegraphics[width=8cm]{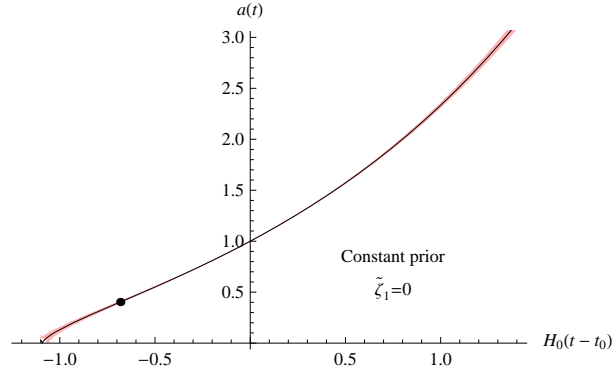}\caption{Plot of the scale factor $a(t)$ with respect to $H_0(t-t_0)$ [see expression (\ref{ScaleFactorExpression})] evaluated at the best estimated value $\tilde{\zeta}_0 = 1.98 \pm 0.06$ (see table \ref{tableSummary_Z01}). It corresponds to the case when $\tilde{\zeta}_1$ is set to zero and it is computed the best estimate for $\tilde{\zeta}_0$ alone, using a constant prior to marginalize over $H_0$. The legend ``Constant prior, $\tilde{\zeta}_1=0$'' indicates this. The plot shows how the scale factor, evaluated at the best estimated value for $\tilde{\zeta}_0$ (with $\tilde{\zeta}_1=0$), predicts a \textit{decelerated} expanding epoch just after de Big-Bang followed by an \textit{accelerating} epoch that is going to continue forever. The transition between deceleration-acceleration happens when the scale factor has a value of $a_t=0.4$ [see table \ref{tableAgeTransition} and expression (\ref{DefScaleFactorTransition})], it is indicated with the black point in the plot. The thin pink band corresponds to evaluate the expression (\ref{ScaleFactorExpression}) at the statistical error of the estimation of $\tilde{\zeta}_0$.}
\label{PlotScaleFactorBestU10Constantz1eq0}
\end{center}
\end{figure}

Now, for the second case $(0< \tilde{\zeta}_0 <3, \; \tilde{\zeta}_1 = 0)$, it has also the same behavior than the case $\tilde{\zeta}_0 >0, \, \tilde{\zeta}_1 <0$ (with $\tilde{\zeta}_0 + \tilde{\zeta}_1 < 3$), it is, the model predicts a Big-Bang as origin of the Universe, followed by an early time decelerated expanding epoch, with a transition to an \textit{accelerating} epoch that is going to continue forever (see figures  \ref{PlotScaleFactorBest12Derive} and \ref{PlotScaleFactorBestU10Constantz1eq0}).
When we set $\tilde{\zeta}_1=0$, the best computed value for the other viscous coefficient is $\tilde{\zeta}_0 = 1.9836 \pm 0.066$ (see table \ref{tableSummary_Z01}).
The transition between deceleration-acceleration happens when the scale factor has a value of $a_t=0.4$ [see table \ref{tableAgeTransition} and expression (\ref{DefScaleFactorTransition})].
In this case  best estimated value for $\tilde{\zeta}_0$ predicts an age for the Universe of 14.82 Gyr, that is in perfect agreement with the constraints of globular clusters.
An important feature of this case is that it does not violate the second law of the thermodynamics (see section \ref{Sectionthermodynamics}).
This particular case has been also studied in \cite{ArturoUlisesPaper1}.

    \section{The curvature}
\label{SectionCurvature}

To study the possible singularities of the model we analyze the behavior of the curvature scalar $R$, as well as the contractions of the Riemann tensor $R^{\alpha \beta \gamma \delta}R_{\alpha \beta \gamma \delta}$ and the Ricci tensor $R^{\alpha \beta}R_{\alpha \beta}$. 
We find that there is not any particular singularity other than that of the Big-Bang, that could indicate possible issues for the model coming from the curvature.
About Weyl tensor, it turns out that  $C^{\alpha \beta \gamma \delta}C_{\alpha \beta \gamma \delta} = 0$ in general for the FRW metric. The details of the analysis is shown below.

    \subsection{The curvature scalar $R$}\label{SectionCurvatureScalar}

The curvature scalar $R \equiv R^{\alpha}_{\;\; \alpha}$ for a spatially flat Universe is given as

\begin{equation}\label{DefScalarCurvature}
R = 6 \left[\frac{\ddot{a}}{a} + H^2\right]
\end{equation}

\noindent The term $\ddot{a}/a$ corresponds to the second Friedmann equation, i.e.,

\begin{equation}\label{2ndFriedmanEq}
\frac{\ddot{a}}{a} = -\frac{4\pi G}{3}(\rho_{\rm m} +3 P^*_{\rm m})
\end{equation}

\noindent We substitute the expressions for: $P^*_{\rm m}=-3\zeta H$ [see eq.
(\ref{pressure})], $\zeta=\zeta_0 + \zeta_1 H$ and $\rho_{\rm m} =
(3/8\pi G) H^2$ from the first Friedmann equation at the expression
(\ref{2ndFriedmanEq}) to obtain

\begin{equation}
\frac{\ddot{a}}{a}= H^2 \left[(12\pi G) \left(\frac{\zeta_0}{H} +
\zeta_1 \right) -\frac{1}{2} \right]
\end{equation}

Using the expressions of the dimensionless bulk viscous coefficients
$\tilde{\zeta}_0$ and $\tilde{\zeta}_1$ [see expressions
(\ref{DefDimensionlessZetas})] yield

\begin{equation}\label{SecondFriedmannEq}
\frac{\ddot{a}}{a} = \frac{H^2}{2}
\left[\frac{H_0}{H}\tilde{\zeta}_0 + \tilde{\zeta}_1 -1\right]
\end{equation}

\noindent We substitute the equation  (\ref{SecondFriedmannEq}) at
(\ref{DefScalarCurvature}) to obtain

\begin{equation}\label{CurvatureScalar2}
R = 3H \left[ H_0 \tilde{\zeta}_0 + H(\tilde{\zeta}_1 +1)
\right]
\end{equation}

\noindent Now, using the equation (\ref{HubbleParameterInTermsOfa}) for the Hubble parameter we arrive to

\begin{equation}\label{CurvatureScalarz1Neq3}
\fl R(a) = \frac{3H_0^2}{(3-\tilde{\zeta}_1)^2} \left\{ 4
\tilde{\zeta}_0^2 + (\tilde{\zeta}_0 + \tilde{\zeta}_1 -3)
a^{(\tilde{\zeta}_1 - 3)/2} \left[(\tilde{\zeta}_1 + 1)
(\tilde{\zeta}_0 + \tilde{\zeta}_1 -3) a^{(\tilde{\zeta}_1 - 3)/2} -\tilde{\zeta}_0
(\tilde{\zeta}_1 + 5)
\right] \right\}
\end{equation}

\noindent for $\tilde{\zeta}_1 \neq 3$.

\begin{figure}
\begin{center}
\includegraphics[width=10cm]{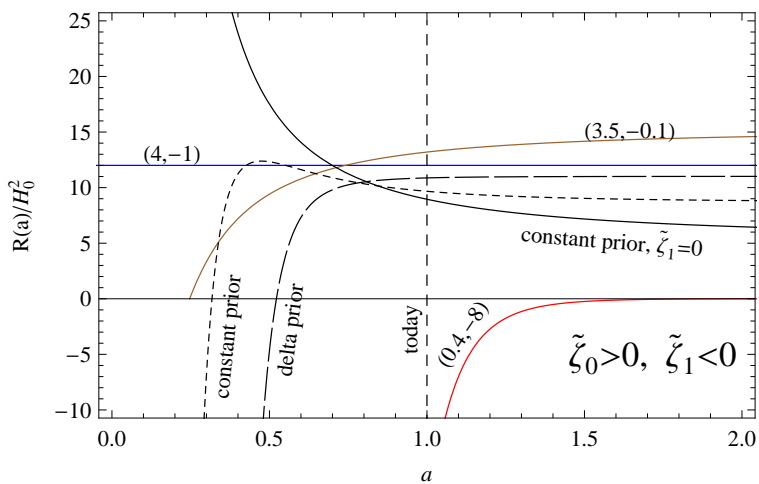}
\caption{Plot of the curvature scalar $R$ with respect to the scale factor for different values of $(\tilde{\zeta}_0, \tilde{\zeta}_1)$
for the cases $(\tilde{\zeta}_0>0,\,\tilde{\zeta}_1<0)$ and $(0<\tilde{\zeta}_0<3,\,\tilde{\zeta}_1=0)$ [see equation
(\ref{CurvatureScalarz1Neq3})]. The short and long dashed lines correspond to evaluate $R$ at the best estimates for $(\tilde{\zeta}_0, \tilde{\zeta}_1)$ when the constant and Dirac delta priors are used to marginalize over $H_0$, respectively.
The solid black line corresponds to evaluate $R$ at the best estimate of $\tilde{\zeta}_0$ alone, when we set $\tilde{\zeta}_1=0$ and marginalize $H_0$ using a constant prior. In general, the curvature scalar diverges when the scale factor is equal to zero for the subcase $\tilde{\zeta}_0 + \tilde{\zeta}_1 < 3$ (with $\tilde{\zeta}_0>0,\,\tilde{\zeta}_1<0$). 
For the subcase $\tilde{\zeta}_0 + \tilde{\zeta}_1 > 3$ (with $\tilde{\zeta}_0>0,\,\tilde{\zeta}_1<0$) the \emph{minimum} value that the scale factor reaches is $a_{\rm min}=[1 + (\tilde{\zeta}_1 - 3)/\tilde{\zeta}_0 ]^{2/(3 - \tilde{\zeta}_1)}$ when $t \rightarrow -\infty$. So, for this subcase the
curvature scalar does not diverge but it is zero when the scale factor is equal to $a_{\rm min}$. Then, there is not the conditions for a Big-Bang.
When $\tilde{\zeta}_0 + \tilde{\zeta}_1 = 3$ the curvature scalar has the \textit{constant value} of $12H^2_0$ along the whole evolution of the Universe. When
$\tilde{\zeta}_1 > -1$ the curvature scalar is \textit{positive} as
$a \rightarrow 0$ and \emph{negative} when $\tilde{\zeta}_1 < -1$.
When $\tilde{\zeta}_1 = -1$ then the curvature scalar is
\textit{positive} as $a \rightarrow 0$ if $\tilde{\zeta}_0 < 4$ and
\emph{negative} if $\tilde{\zeta}_0 > 4$. When $a \rightarrow
\infty$ then $R \rightarrow 12[H_0\tilde{\zeta}_0/(3 -
\tilde{\zeta}_1)]^2$. Finally, for the case $(0<\tilde{\zeta}_0<3,\,\tilde{\zeta}_1=0)$,  the curvature scalar diverges when $a=0$ (the Big-Bang singularity), and when $a \rightarrow \infty$ then $R=\frac{4}{3}H_0^2\tilde{\zeta}_0$.} \label{PlotRZ0Gr0Z1Ls0}
\end{center}
\end{figure}

For the case $\tilde{\zeta}_0 >0, \; \tilde{\zeta}_1<0$, when $a \rightarrow \infty$ then $R \rightarrow 12[H_0\tilde{\zeta}_0 / (3 -\tilde{\zeta}_1)]^2$. When in addition $\tilde{\zeta}_0 + \tilde{\zeta}_1 < 3$, then the curvature scalar diverges when the scale factor $a$ is equal to zero. This supports the existence of a Big-Bang in the past
of the Universe (see figure \ref{PlotRZ0Gr0Z1Ls0} and section \ref{SectionScaleFactor}).

However, when $\tilde{\zeta}_0 + \tilde{\zeta}_1 > 3$ (with $\tilde{\zeta}_0 >0, \; \tilde{\zeta}_1<0$) there is not Big-Bang in the past of the Universe because the scale factor never becomes zero in the past,
the \emph{minimum} value that the scale factor can reach is $a_{\rm min}=[1+ (\tilde{\zeta}_1-3)/\tilde{\zeta}_0
]^{2/(3-\tilde{\zeta}_1)}$ when $t \rightarrow -\infty$. So, the curvature scalar does not diverge but it is zero when the scale factor is equal to $a_{\rm min}$. Then, there is not the conditions for a Big-Bang when $\tilde{\zeta}_0 + \tilde{\zeta}_1 > 3$ (see figure \ref{PlotRZ0Gr0Z1Ls0}).

When $\tilde{\zeta}_0 + \tilde{\zeta}_1 = 3$ the curvature scalar has the \textit{constant value} of $12H^2_0$ along the whole evolution of the Universe.
When $\tilde{\zeta}_1 > -1$ the curvature scalar is
\textit{positive} as $a \rightarrow 0$ and \emph{negative} when $\tilde{\zeta}_1 < -1$. When $\tilde{\zeta}_1 = -1$ then the curvature scalar is \textit{positive} as $a \rightarrow 0$ if $\tilde{\zeta}_0 < 4$ and \emph{negative} if $\tilde{\zeta}_0 > 4$.
The present central value of the curvature scalar is $R/H_0^2=9.66$ and  $R/H_0^2=10.87$ when constant and Dirac delta priors for $H_0$ are used, respectively.

For the other important case $(0<\tilde{\zeta}_0<3, \, \tilde{\zeta}_1=0)$, we find that the curvature scalar diverges when $a \rightarrow 0$ (the Big-Bang singularity). From the infinite value of $R(a=0)$ it decreases as the Universe expands until to reach its minimum value of $R=\frac{4}{3}H_0^2\tilde{\zeta}_0$ when $a \rightarrow \infty$ (see figure \ref{PlotRZ0Gr0Z1Ls0}).

    \subsection{The scalar $R^{\alpha \beta \gamma \delta}R_{\alpha \beta \gamma \delta}$}

We analyze the behavior of the contraction of the Riemann tensor with itself, i.e., $K\equiv R^{\alpha \beta \gamma \delta}R_{\alpha \beta \gamma \delta}$.
In general, for a FRW metric in a curved space-time $K$ is found to be

\begin{equation}\label{DefKRiemannGeneralDef}
K \equiv R^{\alpha \beta \gamma \delta}R_{\alpha \beta \gamma \delta} = 12 \left[\left(\frac{\dot{a}}{a} \right)^4 + 2k  \left(\frac{\dot{a}}{a^2} \right)^2 + \frac{k^2}{a^4} + \left(\frac{\ddot{a}}{a} \right)^2 \right]
\end{equation}

\noindent where $k$ characterizes the constant \textit{spatial} curvature of the FRW metric; for our model $k=0$. So, using the fact of that $H \equiv (\dot{a}/a)$ and that $\ddot{a}/a$ corresponds to the second Friedmann equation given by expression (\ref{SecondFriedmannEq}) we rewrite the equation (\ref{DefKRiemannGeneralDef}) as

\begin{equation}
K  = 12 H^4 \left[1 + \frac{1}{4} \left(\frac{H_0 \tilde{\zeta}_0}{H} + \tilde{\zeta}_1 -1 \right)^2 \right]
\end{equation}

\noindent where  $H$ is given by the expression (\ref{HubbleParameterInTermsOfa}), that we substitute to obtain

\begin{eqnarray}\label{KRiemann}
\fl K(a)= \frac{12H_0^4}{(\tilde{\zeta}_1-3)^4 \, a^6} \left\{ \tilde{\zeta}_0 a^{3/2} + (3-\tilde{\zeta}_0-\tilde{\zeta}_1) a^{\tilde{\zeta}_1/2} \right\}^4 \times \\
\times \left\{1 + \frac{1}{4}\left[1-\tilde{\zeta}_1 + \frac{\tilde{\zeta}_0(\tilde{\zeta}_1-3) a^{3/2}}{\tilde{\zeta}_0 a^{3/2} + (3-\tilde{\zeta}_0-\tilde{\zeta}_1)a^{\tilde{\zeta}_1/2}} \right]^2 \right\} \nonumber
\end{eqnarray}

\begin{figure}
\begin{center}
\includegraphics[width=16cm]{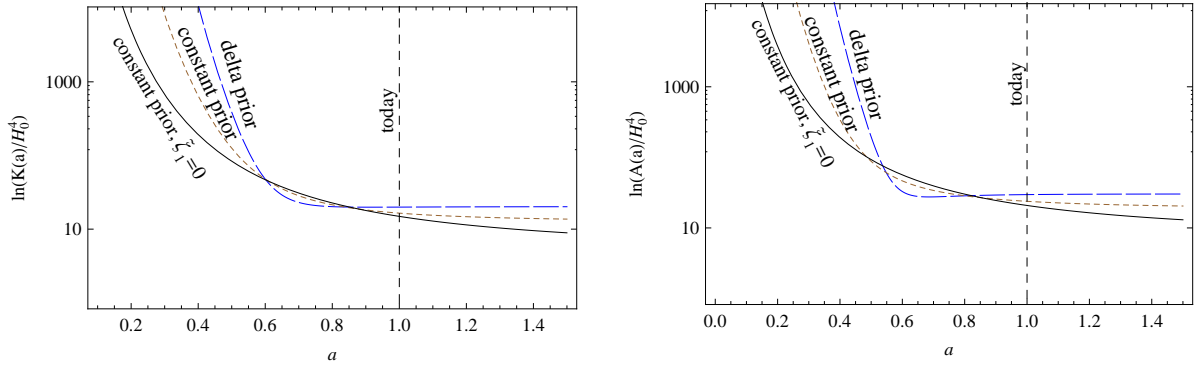}
\caption{Plot in logarithmic scale  of the scalars of curvature $K$ (left) and $A$ (right), coming from the contraction of the Riemann tensor $K \equiv R^{\alpha \beta \gamma \delta}R_{\alpha \beta \gamma \delta}$, and the Ricci tensor $A\equiv~R^{\alpha \beta}R_{\alpha \beta}$,  with respect to the scale factor [see expressions (\ref{KRiemann}) and (\ref{ARicciContraction})  respectively]. Both scalars are evaluated at the best estimates for $(\tilde{\zeta}_0, \tilde{\zeta}_1)$ together, and $\tilde{\zeta}_0$ alone (when we set $\tilde{\zeta}_1=0$). See table \ref{tableSummary_Z01}. Note that in both figures and for the six plots, $K$ and $A$ diverge when $a \rightarrow 0$ (the Big-Bang singularity) and when $a \rightarrow \infty$ then $K$ and $A$ go to a constant minimum value.}
\label{PlotRiemannRicci}
\end{center}
\end{figure}

For the case $(\tilde{\zeta}_0 >0, \, \tilde{\zeta}_1<0)$, we find that $K$ diverges to infinity when $a \rightarrow 0$ and when $a \rightarrow \infty$ then $K \rightarrow 24[H_0 \tilde{\zeta}_0/(\tilde{\zeta}_1-3)]^4$. For the best estimates, the present central value for the Riemann contraction is $K/H_0^4=16.48$ and $K/H_0^4=19.91$ when constant and Dirac delta priors are used, respectively.
Figure \ref{PlotRiemannRicci} shows the behavior of $K$ with respect to the scale factor, evaluated at the best estimates.

For the other case of interest $(0<\tilde{\zeta}_0<3, \,\tilde{\zeta}_1=0)$, we find that the scalar $K$ is infinite when $a \rightarrow 0$ (the Big-Bang singularity), it decreases from $K(a=0)$ as the Universe expands until to reach its minimum value of $K=(8/27)(H_0 \tilde{\zeta}_0)^4$ when $a \rightarrow \infty$ (see figure \ref{PlotRiemannRicci}).

    \subsection{The scalar $R^{\alpha \beta}R_{\alpha \beta}$}

The scalar derived of contracting the Ricci tensor with itself,  $A \equiv R^{\alpha \beta}R_{\alpha \beta}$, for a curved FRW metric has the form

\begin{equation}
\fl A \equiv R^{\alpha \beta} R_{\alpha \beta} = 12 \left[ \left(\frac{\ddot{a}}{a} \right)^2 + \frac{\ddot{a}}{a^3}(\dot{a}^2 + k)  + \left(\frac{\dot{a}}{a}\right)^4 + 2k \left(\frac{\dot{a}}{a^2} \right)^2 + \left(\frac{k}{a^2} \right)^2 \right]
\end{equation}

\noindent For our model $k=0$. We substitute $\dot{a}/a$ by $H$ and  $\ddot{a}/a$ by the expression (\ref{SecondFriedmannEq}) to obtain

\begin{equation}\label{AWithk0}
A = 3H^2 \left[H_0^2 \tilde{\zeta}_0^2 + 2 H_0 \tilde{\zeta}_0 \tilde{\zeta}_1 H + (3+\tilde{\zeta}_1^2)H^2  \right]
\end{equation}

\noindent We put the expression for the Hubble constant (\ref{HubbleParameterInTermsOfa}) in (\ref{AWithk0}) yielding

\begin{eqnarray}\label{ARicciContraction}
\fl A(a)=\frac{3H_0^4}{(\tilde{\zeta}_1-3)^4 a^6} \left\{ \tilde{\zeta}_0 a^{3/2} + (3-\tilde{\zeta}_0-\tilde{\zeta}_1)a^{\tilde{\zeta}_1/2} \right\}^2 \times \\
\times \left\{12\tilde{\zeta}_0^2 a^3 + 6 \tilde{\zeta}_0 (1+ \tilde{\zeta}_1)(3-\tilde{\zeta}_0 - \tilde{\zeta}_1) a^{(3-\tilde{\zeta}_1)/2}- (3+\tilde{\zeta}_1^2)(3-\tilde{\zeta}_0 -\tilde{\zeta}_1)^2 a^{\tilde{\zeta}_1}  \right\} \nonumber
\end{eqnarray}

For the case $(\tilde{\zeta}_0 >0, \tilde{\zeta}_1<0)$,  when $a \rightarrow 0$ then $A \rightarrow \infty$. And  when $a \rightarrow \infty$ then $A \rightarrow 36[H_0 \tilde{\zeta}_0/(\tilde{\zeta}_1-3)]^4$. For the best estimates, the present central value of the Ricci contraction is $A/H_0^4=23.81$ and $A/H_0^4=29.66$ when constant and Dirac delta priors are used, respectively.
Figure \ref{PlotRiemannRicci} shows the behavior of $A$ with respect to the scale factor.

For the other case  $(0<\tilde{\zeta}_0<3, \, \tilde{\zeta}_1=0)$, the scalar $A$ is infinite when $a \rightarrow 0$ (the Big-Bang singularity), it decreases from $A(a=0)$ as the Universe expands until to reach its minimum value of $A=(4/9)(H_0 \tilde{\zeta}_0)^4$ when $a \rightarrow \infty$ (see figure \ref{PlotRiemannRicci}).

    \section{Matter density $\rho_{\rm m}$} \label{SectionMatterDensity}

We analyze also the equation of the matter density to have a better understanding of the behavior of the Universe according to the model. From the first Friedmann equation~(\ref{FriedmannFirstEquation}) we have

\begin{equation}\label{DensityMatterBase}
\rho_{\rm m}(a) = \left( \frac{3}{8\pi G} \right) H^2(a)
\end{equation}

\noindent Using expression for the Hubble parameter
(\ref{HubbleParameterInTermsOfa}) at the expression above we arrive
to

\begin{equation}\label{DensityMatterGeneral}
\rho_{\rm m}(a) = \rho^0_{\rm crit} \left[\left( 1 -
\frac{\tilde{\zeta}_0}{3-\tilde{\zeta}_1} \right)
a^{(\tilde{\zeta}_1 -3 )/2}
+\frac{\tilde{\zeta}_0}{3-\tilde{\zeta}_1} \right]^2; \quad
{\mbox{for}} \; \tilde{\zeta}_1 \neq 3
\end{equation}

\begin{figure}
\begin{center}
\includegraphics[width=16cm]{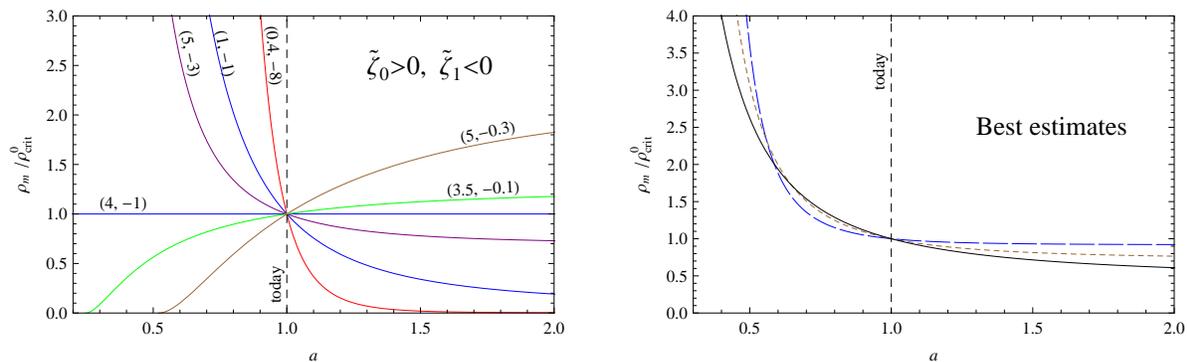}
\caption{Plot of the matter density $\rho_{\rm m}$ with respect to the scale factor for different values of $(\tilde{\zeta}_0,
\tilde{\zeta}_1)$ where $\tilde{\zeta}_0 > 0$ and $\tilde{\zeta}_1 < 0$ (left panel) and also in the best estimates for $(\tilde{\zeta}_0,\tilde{\zeta}_1)$ (right panel) [see equation (\ref{DensityMatterGeneral}) and table \ref{tableSummary_Z01}].
For the subcase   $\tilde{\zeta}_0 + \tilde{\zeta}_1 < 3$ (with $\tilde{\zeta}_0 > 0, \tilde{\zeta}_1<0$) the matter density diverges when the scale factor is equal to zero. This supports the assumption of a Big-Bang as the beginning of the Universe (see section \ref{SectionScaleFactor}). However, for the subcase $\tilde{\zeta}_0 + \tilde{\zeta}_1 > 3$ the minimum value that the scale factor reaches is $a_{\rm min}=[(\tilde{\zeta}_0 + \tilde{\zeta}_1
-3)/\tilde{\zeta}_0]^{2/(3-\tilde{\zeta}_1)}$ when  $t \rightarrow -
\infty$. The matter density is zero when the scale factor is equal to $a_{\rm min}$. Then, there is not the conditions for a Big-Bang in this subcase.
In both subcases, when $a \rightarrow \infty$ then $\rho_{\rm m} = \rho^0_{\rm crit} [\tilde{\zeta}_0/(3 - \tilde{\zeta}_1)]^2$.
For the subcase $\tilde{\zeta}_0 + \tilde{\zeta}_1 = 3$, the matter density has the \textit{constant} value of $\rho^0_{\rm crit}$. For the right panel, the short and long dashed lines correspond to evaluate the matter density at the best estimates when it is used the constant and Dirac delta priors to marginalize over $H_0$, respectively. The solid black line corresponds to the best estimate for $\tilde{\zeta}_0$, setting $\tilde{\zeta}_1=0$ and using a constant prior to marginalize over $H_0$.}
\label{PlotRhoZ0Gr0Z1Ls0}
\end{center}
\end{figure}

When $\tilde{\zeta}_0 > 0, \; \tilde{\zeta}_1<0$: For the subcase   $\tilde{\zeta}_0 + \tilde{\zeta}_1 < 3$ (with $\tilde{\zeta}_0 > 0, \tilde{\zeta}_1<0$) the matter density diverges when the scale factor is equal to zero. This supports the assumption of a Big-Bang as the beginning of the Universe (see section \ref{SectionScaleFactor} and figure~\ref{PlotRhoZ0Gr0Z1Ls0}).

However, for the subcase $\tilde{\zeta}_0 + \tilde{\zeta}_1 > 3$ the minimum value that the scale factor reaches is $a_{\rm min}=[(\tilde{\zeta}_0 + \tilde{\zeta}_1
-3)/\tilde{\zeta}_0]^{2/(3-\tilde{\zeta}_1)}$ when  $t \rightarrow -
\infty$. The matter density is zero when the scale factor is equal to $a_{\rm min}$. Then, there is not the conditions for a Big-Bang in this subcase.
In both subcases, when $a \rightarrow \infty$ then $\rho_{\rm m} = \rho^0_{\rm crit} [\tilde{\zeta}_0/(3 - \tilde{\zeta}_1)]^2$.

For the subcase $\tilde{\zeta}_0 + \tilde{\zeta}_1 = 3$, the matter density has the \textit{constant} value of $\rho^0_{\rm crit}$. Figure \ref{PlotRhoZ0Gr0Z1Ls0} shows the behavior of the matter density with respect to the scale factor.

For the other case of interest, $(0<\tilde{\zeta}_0<3,\,\tilde{\zeta}_1=0)$,  the matter density is infinite when $a \rightarrow 0$ (the Big-Bang singularity), from this infinite value $\rho_{\rm m}$ decreases as the Universe expands until to reach its minimum value of $\rho_{\rm m}= (H_0^2/24 \pi G) \tilde{\zeta}_0^2$ when $a \rightarrow \infty$ (see figure \ref{PlotRhoZ0Gr0Z1Ls0}).

\begin{figure}
\begin{center}
\includegraphics[width=16cm]{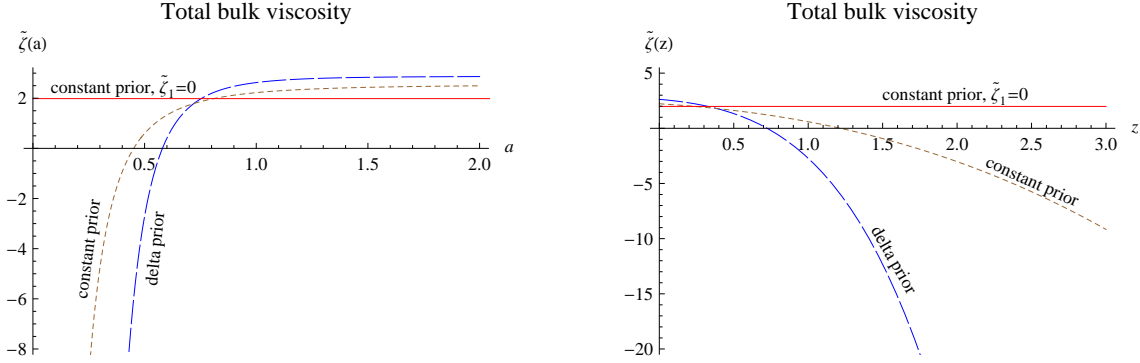}
\caption{Plot of the total dimensionless bulk viscosity $\tilde{\zeta}= \tilde{\zeta}_0 + \tilde{\zeta}_1 H$ with respect to the scale factor (left panel) and the redshift (right panel) [see equation (\ref{TotalDimensionlessViscosity})] when it is evaluated at the best estimated values for $(\tilde{\zeta}_0,\tilde{\zeta}_1)$  (see table \ref{tableSummary_Z01}). At the early Universe (small values of $a$, or greater values of $z$) the \textit{total} bulk viscosity $\tilde{\zeta}$ is negative for the case of ``delta prior'' and ``constant prior''. Negative values for the total bulk viscosity $\tilde{\zeta}$ implies a violation of the local second law of thermodynamics at early times [see equation (\ref{entropy_condition})]. So, the  model, $\tilde{\zeta}=\tilde{\zeta}_0+\tilde{\zeta}_1H$,  works well only for recent times of the Universe. The plot ``constant prior, $\tilde{\zeta}_1=0$'' corresponds to the model $\tilde{\zeta}=\tilde{\zeta}_0$ evaluated at the best estimated value for $\tilde{\zeta}_0$ (we have set $\tilde{\zeta}_1=0$). This case does not violate the local second law of thermodynamics, the supernova data indicates a positive value for $\tilde{\zeta}_0$ and it is constant along the whole history of the Universe. The transition to a positive values in recent times of the model $\tilde{\zeta}=\tilde{\zeta}_0+\tilde{\zeta}_1H$ happens when the value of the scale factor $a_{\rm np}$ is  given by equation (\ref{ScaleFactorViscosityTransitionNegativePositive}). And when $a \rightarrow \infty$,  the total bulk viscosity for this case is $\tilde{\zeta} = 3\tilde{\zeta}_0/(3 - \tilde{\zeta}_1)$.}
\label{PlotTotalViscosityU10SNeOnlyAZ}
\end{center}
\end{figure}

        \section{Thermodynamics and the local entropy.}
        \label{Sectionthermodynamics}

The law of generation of \textit{local} entropy in a fluid on a FRW
space-time can be written as \cite{Weinberg}

\begin{equation}\label{entropy_definition}
T \, \nabla_{\nu} s^{\nu} = \zeta (\nabla_{\nu} u^{\nu})^2 = 9H^2 \zeta
\end{equation}

\noindent where $T$ is the temperature and $\nabla_{\nu} s^{\nu} $
is the rate of entropy production in a unit volume. With this, the
second law of the thermodynamics can be written as

\begin{equation}\label{2ndLawThermodynamics}
 T \nabla_{\nu} s^{\nu} \geq 0
\end{equation}

\noindent so, from the expression \eref{entropy_definition}, it implies that

\begin{equation}\label{2ndLawThermodynamicsZeta}
\zeta \geq 0
\end{equation}

\noindent Thus, for the present model the inequality (\ref{2ndLawThermodynamicsZeta}) can be written as

\begin{equation}\label{entropy_condition}
\zeta =  \zeta_0 + \zeta_1 H \;  \geq 0
\end{equation}

\noindent Using the expression for the Hubble parameter (\ref{HubbleParameterInTermsOfa}) we find that the expression for the \textit{total} bulk viscosity $\zeta(a)$ is given as

\begin{equation}\label{TotalDimensionlessViscosity}
\tilde{\zeta}(a)= \tilde{\zeta}_0 + \tilde{\zeta}_1 \left[ \left(1 - \frac{\tilde{\zeta}_0}{3 - \tilde{\zeta}_1} \right)
a^{(\tilde{\zeta}_1-3)/2} + \frac{\tilde{\zeta}_0}{3-\tilde{\zeta}_1} \right]
\end{equation}

\noindent where we have defined the  \textit{total dimensionless} bulk viscous coefficient $\tilde{\zeta} \equiv (24 \pi G / H_0) \zeta$. Figure \ref{PlotTotalViscosityU10SNeOnlyAZ} shows the behavior of $\tilde{\zeta}$ with respect to the scale factor when $(\tilde{\zeta}_0, \tilde{\zeta}_1)$ are evaluated at the best estimated value (see section \ref{SectionSNeTest} and table \ref{tableSummary_Z01}).
We find that for the best estimates of $(\tilde{\zeta}_0, \tilde{\zeta}_1)$, the \textit{total} bulk viscosity is \textit{negative} in early times of the Universe and \textit{positive} at late times.
For the best estimate of $\tilde{\zeta}_0$ alone (when we set $\tilde{\zeta}_1=0$), the total bulk viscosity is \textit{positive} and constant along the cosmic time, because in this case $\tilde{\zeta}=\tilde{\zeta}_0$.

According to equation (\ref{entropy_condition}), negative values for the total bulk viscosity $\tilde{\zeta}$ implies a violation of the local second law of thermodynamics. So, the model $\tilde{\zeta}=\tilde{\zeta}_0+\tilde{\zeta}_1H$, evaluated at the best estimates, violates the entropy law at early times, but the model $\tilde{\zeta}=\tilde{\zeta}_0$, evaluated at the best estimate for $\tilde{\zeta}_0$ (with $\tilde{\zeta}_1=0$), does not.

The \textit{transition} between negative to positive values of the total bulk viscosity for the case $\tilde{\zeta}=\tilde{\zeta}_0+\tilde{\zeta}_1H$ happens when the scale factor has the value

\begin{equation}\label{ScaleFactorViscosityTransitionNegativePositive}
a_{\rm np} = \left[\frac{3\tilde{\zeta}_0}{\tilde{\zeta}_1(\tilde{\zeta}_0+\tilde{\zeta}_1-3)} \right]^{2/(\tilde{\zeta}_1-3)}
\end{equation}

\noindent The subscript ``np'' stands for ``negative to positive'' values. For the best estimates for $(\tilde{\zeta}_0,\tilde{\zeta}_1)$ using the SNe Ia probe we find that the value of the scale factor $a_{\rm np}$ when transition happens are $a_{\rm np}=0.45$ and $a_{\rm np}=0.57$  when the \textit{constant} and \textit{Dirac delta}  priors are assumed to marginalize over $H_0$, respectively.
When $a \rightarrow \infty$ then the total bulk viscosity is $\tilde{\zeta} = 3\tilde{\zeta}_0/(3-\tilde{\zeta}_1)$.

                \section{Type Ia Supernovae test.}
                \label{SectionSNeTest}

\begin{table}
  \centering
\begin{tabular}{| c  | c c | c c |}
\multicolumn{5}{c}{\textbf{Viscous model $\zeta=\zeta_0 + \zeta_1 H$}}\\
\hline

Prior over $H_0$ &   $\tilde{\zeta}_0$ & $\tilde{\zeta}_1$ & $\chi^2_{{\rm min}}$ & $\chi^2_{{\rm d.o.f.}}$ \\
\hline

Constant &  $4.38929 \pm 1.568$ & $-2.16673 \pm 1.421$ & 542.38 & 0.977 \\
Dirac delta &  $8.58935 \pm 1.217$ & $-5.96505 \pm 1.159$ & 561.15 & 1.011 \\
Constant ($\tilde{\zeta}_1=0$) &  $1.9835 \pm 0.066$ & 0 & 544.587 & 0.979 \\
\hline

\end{tabular}
\caption{Summary of the best estimated values of the dimensionless coefficients  $\tilde{\zeta}_0$ and $\tilde{\zeta}_1$ for a bulk viscous matter-dominated  model with a bulk viscosity of the form $\zeta = \zeta_0 + \zeta_1 H$.
The best estimates were computed minimizing a $\chi^2$ function of a Bayesian statistical analysis using the SCP ``Union2'' 2010 compilation data set composed by 557 type Ia SNe \cite{AmanullahUnion22010}.
The first column shows the prior distribution function assumed to marginalize over the nuisance parameter $H_0$ and then compute the best estimates for $\tilde{\zeta}_0$ and $\tilde{\zeta}_1$ once $H_0$ has been marginalized. There are several ways to marginalize over $H_0$, in the present work we choose two:  (1) \textit{``Constant prior''} means that we do not assume any particular value for $H_0$, therefore, its probability distribution function is \textit{constant}. (2) \textit{``Dirac delta prior''} means that we \textit{do} assume a \textit{specific} value for $H_0$ (in particular we used $H_0= 72 \; ({{\rm km}}/{{\rm s}}){{\rm Mpc}}^{-1}$), therefore, its probability distribution function has the form of a \textit{Dirac delta} centered at $H_0= 72 \; ({{\rm km}}/{{\rm s}}){{\rm Mpc}}^{-1}$.
In the third row, the label  ``Constant ($\tilde{\zeta}_1=0$)'' means that we set $\tilde{\zeta}_1=0$ and use a \textit{constant} prior for $H_0$, to compute the best estimated value for $\tilde{\zeta}_0$ alone.
The second column corresponds to the best estimated values computed for $(\tilde{\zeta}_0, \tilde{\zeta}_1)$, or $\tilde{\zeta}_0$ alone (third row). The third column shows $\chi^2_{\rm min}$, the minimum value obtained for the $\chi^2$ function (\ref{ChiSquareDefinition}), and also the $\chi^2$~function by \textit{degrees of freedom}:  $\chi^2_{\rm d.o.f.}$. It is defined as $\chi^2_{\rm d.o.f.} \equiv \chi^2_{\rm min}/(n-p)$ where $n$ is the number of data (in our case $n=557$) and $p$ is the number of free parameters of the model (for the first two raws $p=2$ and for the third one $p=1$).
The errors are at 68.3\% confidence level. Figures
\ref{PlotAllCIZ01SNeU10FDM}  and \ref{PlotAllCIZ01SNeU10Dirac} show the confidence intervals.} \label{tableSummary_Z01}
\end{table}

We test and constrain the viability of the model using the type Ia Supernovae  (SNe Ia) observations. So, we calculate the \textit{best estimated values}
for the parameters $\tilde{\zeta}_0$ and $\tilde{\zeta}_1$ and the
\emph{goodness-of-fit} of the model to the data by a
$\chi^2$-minimization and then compute the confidence intervals for $(\tilde{\zeta}_0, \tilde{\zeta}_1)$ to constrain their possible values with levels of statistical confidence.
We use the ``Union2'' SNe Ia data set (2010) from ``The Supernova Cosmology Project'' (SCP) composed by 557 type Ia supernovae \cite{AmanullahUnion22010}.

\begin{figure}
\begin{center}
\includegraphics[width=10cm]{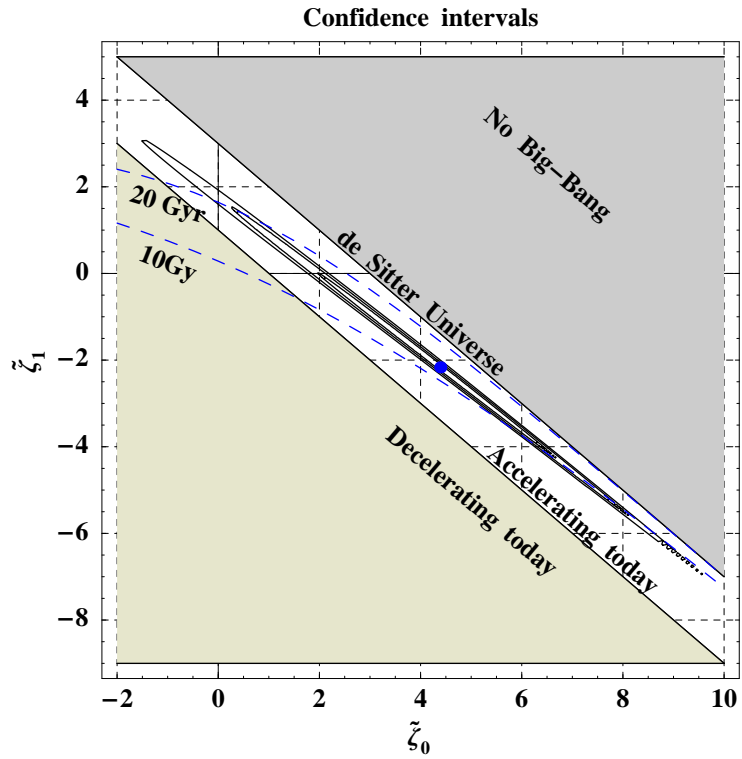}
\caption{Confidence intervals for the dimensionless coefficient
$(\tilde{\zeta}_0,\tilde{\zeta}_1)$ of a bulk viscous
matter-dominated Universe model with bulk viscosity of the form
$\zeta=\zeta_0 + \zeta_1 H$.  We used the SCP (2010) ``Union2'' SNe data set composed of 557 type Ia SNe \cite{AmanullahUnion22010}, to compute the best estimates $\tilde{\zeta}_0=4.389 \pm 1.56$ and $\tilde{\zeta}_1=-2.166 \pm 1.42$; they are indicated with the blue point in the figure. The confidence intervals shown correspond to $68.3\% $, $95.4\%$ and $99.73\%$ of probability. The grey region (upper region) corresponds to the values for 
$(\tilde{\zeta}_0,\tilde{\zeta}_1)$ to which the model $\zeta=\zeta_0 + \zeta_1 H$ predicts that there is not a Big-Bang in the past, it is, the scale factor never becomes zero in the past of the Universe (see figure \ref{PlotScaleFactorZ0Gr0Z1Ls0Gr3}), as well as the curvature scalar $R$ did not diverge in the past (see figure \ref{PlotRZ0Gr0Z1Ls0}). The boundary line between the grey and white regions corresponds to a \textit{de Sitter Universe}, it is obtained when $\tilde{\zeta}_0+\tilde{\zeta}_1=3$.
The white region (central region) indicates the values to which the Universe is \textit{accelerating} today, and the green region (lower) corresponds to deceleration today. Note that the confidence intervals of the best estimates for $(\tilde{\zeta}_0,\tilde{\zeta}_1)$ lie precisely in the accelerating region today with $99.73 \%$ of probability, at least. The dashed blue lines show two values for the age of the Universe  (10 and 20 Gyr). The computed central value for the age using the best estimates for $(\tilde{\zeta}_0,\tilde{\zeta}_1)$ is 11.72 Gyr (see table \ref{tableAgeTransition}). The Hubble constant $H_0$ was marginalized assuming a \textit{constant prior} distribution.}
\label{PlotAllCIZ01SNeU10FDM}
\end{center}
\end{figure}

We use the definition of luminosity distance $d_L$
(see \cite{Riess2004,Riess2006}, \cite{TurnerRiess}--\cite{Copeland}) in a
flat cosmology,
\begin{eqnarray}\label{luminosity_distance1}
d_L(z, \tilde{\zeta}_0,\tilde{\zeta}_1, H_0) &=& c(1+z) \int_0^z
\frac{dz'}{H(z', \tilde{\zeta}_0,\tilde{\zeta}_1, H_0)}
\end{eqnarray}

\noindent where $H(z, \tilde{\zeta}_0,\tilde{\zeta}_1, H_0)$ is the Hubble parameter, i.e., the expression
(\ref{HubbleParameter}) and `$c$' is the speed of light given in units of km/sec. The \emph{theoretical distance moduli} for the $k$-th supernova with redshift $z_k$ is defined as

\begin{equation}\label{distanceModuli}
\mu^{{\rm t}}(z_k, \tilde{\zeta}_0, \tilde{\zeta}_1, H_0) \equiv m-M = 5\log_{10} \left[\frac{d_L(z_k, \tilde{\zeta}_0, \tilde{\zeta}_1, H_0)}{{\rm Mpc}} \right] +25
\end{equation}

\noindent where $m$ and $M$ are the apparent and absolute magnitudes of the SNe Ia respectively, and the superscript `t' stands for \textit{``theoretical''}. We construct the statistical $\chi^2$ function as

\begin{equation}\label{ChiSquareDefinition}
\chi^2 (\tilde{\zeta}_0, \tilde{\zeta}_1, H_0) \equiv
\sum_{k = 1}^n \frac{\left[\mu^{{\rm t}} (z_k, \tilde{\zeta}_0,
\tilde{\zeta}_1, H_0) - \mu_k \right]^2}{\sigma_k^2}
\end{equation}

\noindent where $\mu_k$ is the \emph{observational} distance moduli for the $k$-th supernova, $\sigma_k^2$ is the variance of the measurement and $n$ is the amount of supernova in the data set, in this case $n=557$.

\begin{figure}
\begin{center}
\includegraphics[width=10cm]{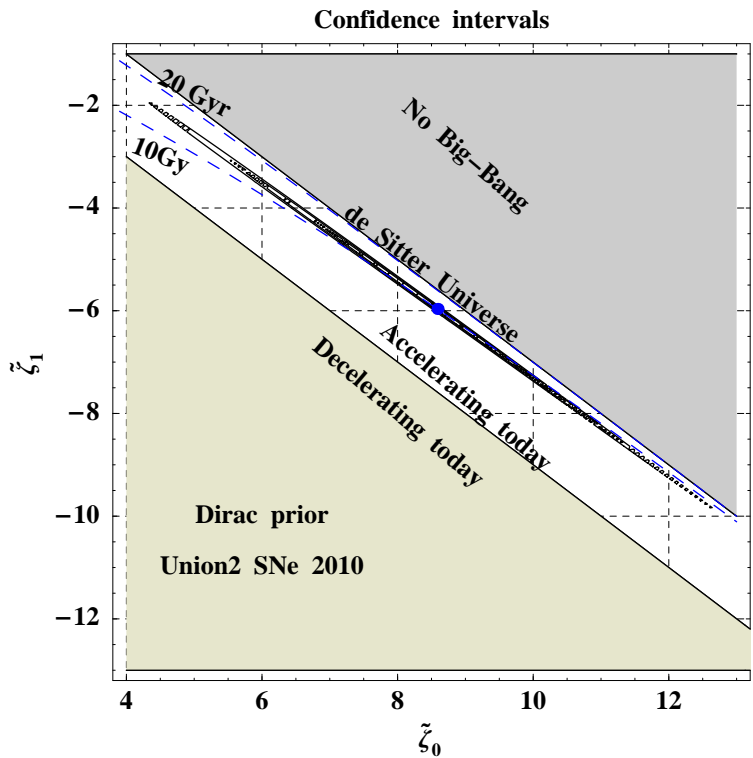}
\caption{Confidence intervals for the dimensionless coefficient
$(\tilde{\zeta}_0,\tilde{\zeta}_1)$ of a bulk viscous
matter-dominated Universe model with bulk viscosity of the form
$\zeta=\zeta_0 + \zeta_1 H$. It is like the figure \ref{PlotAllCIZ01SNeU10FDM} but now a \textit{Dirac delta prior} was used to marginalize over $H_0$.
The best estimates are $\tilde{\zeta}_0= 8.58 \pm  1.2$ and $\tilde{\zeta}_1=-5.96  \pm  1.15$ (see table \ref{tableSummary_Z01}).
Note how the confidence intervals of the best estimates lie precisely in the accelerating region today with $99.73 \%$ of probability. The computed central value for the age using the best estimates  is 10.03 Gyr for this case (see table \ref{tableAgeTransition}).}
\label{PlotAllCIZ01SNeU10Dirac}
\end{center}
\end{figure}

Once constructed the $\chi^2$ function (\ref{ChiSquareDefinition})
we  numerically minimize it to compute the ``\textit{best
estimates}'' for the free parameters $(\tilde{\zeta}_0, \tilde{\zeta}_1)$ of the model. The minimum value of the $\chi^2$ function gives the best estimated values of
$(\tilde{\zeta}_0, \tilde{\zeta}_1)$ and measures the
goodness-of-fit of the model to data. The results are shown in table \ref{tableSummary_Z01} and the confidence intervals for $\tilde{\zeta}_0$ vs $\tilde{\zeta}_1$ are shown in figures \ref{PlotAllCIZ01SNeU10FDM} and \ref{PlotAllCIZ01SNeU10Dirac}.

About the value of the Hubble constant $H_0$ we marginalize it using two methods:
\begin{description}
\item[Constant prior]: We assume that $H_0$ does not have any preferred value a priori, i.e, it has a \textit{constant} prior probability distribution function. So, we solve \textit{analytically} certain integrals using this assumption to obtain a $\chi^2$-function like the expression (\ref{ChiSquareDefinition}) but that it is not going to depend on the variable $H_0$ anymore. This method is carefully described in the appendix A of \cite{ArturoUlisesPaper1}.
\item[Dirac delta prior]: We assume that $H_0$ has a \textit{specific} value (suggested by some other independent observation), so, its probability distribution function has the form of a \textit{Dirac delta}, centered at the specific value. In particular we choose $H_0= 72 \; ({{\rm km}}/{{\rm s}}){{\rm Mpc}}^{-1}$ as suggested by the observations of the Hubble Space Telescope (HST) \cite{Freedman2001}. In practice, it simply means to set $H_0= 72$ at the expression (\ref{ChiSquareDefinition}) [i.e., at expression (\ref{HubbleParameter})] and with this $H_0$ is not a free parameter of the model anymore.
\end{description}

\begin{figure}
\begin{center}
\includegraphics[width=7cm]{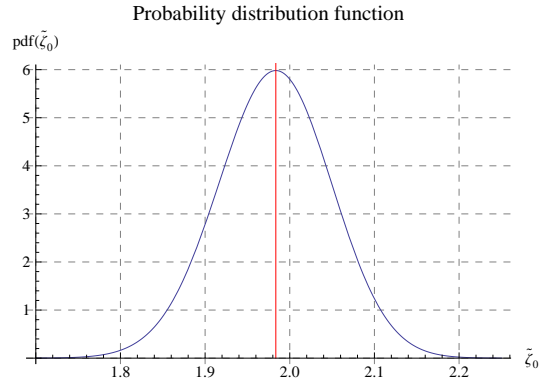}
\caption{Plot of the normalized  Probability Distribution Function (PDF) of $\tilde{\zeta}_0$ for the case when we set $\tilde{\zeta}_1=0$. The computed best estimated value (the central value of the PDF) is $\tilde{\zeta}_0=1.9835 \pm 0.066$, the error is given at $68.3 \%$ confidence level (see table~\ref{tableSummary_Z01}). A constant prior was used to marginalize over $H_0$.}
\label{plotPDFz0FDMU10Final}
\end{center}
\end{figure}

        \section{Conclusions.}
        \label{SectionConclusions}

We have explored a bulk viscous matter-dominated cosmological model composed by a  pressureless fluid (dust) with bulk viscosity of the form $\zeta = \zeta_0 + \zeta_1 H$, representing both baryon and dark matter components.

One of the advantage of this model is that it resolves automatically the ``Coincidence problem'' because it is not introduced any dark energy component, i.e.,  it is enough to think of the Universe as composed by baryon and dark matter components as dominant ingredients of the Cosmos.

The expansion of the Universe can be seen as a collection of states  \textit{out} of the thermal equilibrium during \textit{very short} periods of time where it is produced local entropy.

From  all the possible scenarios predicted by the  model  according to different values of the dimensionless bulk viscous coefficients $\tilde{\zeta}_0$ and $\tilde{\zeta}_1$, we find two  where the Universe begins with a Big-Bang, followed by a decelerated expansion epoch at early times, and with a transition to an accelerated expansion at recent times that is going to continue forever. These two scenarios are characterized for the ranges $(\tilde{\zeta}_0 >0, \tilde{\zeta}_1<0)$ with $\tilde{\zeta}_0 + \tilde{\zeta}_1<3$, and $(0<\tilde{\zeta}_0 <3, \tilde{\zeta}_1=0)$. This behavior of the Universe corresponds to what is actually observed and is close to the $\Lambda$CDM paradigm.

On the other hand, we compute also  the best estimated  values for $(\tilde{\zeta}_0, \tilde{\zeta}_1)$ using the latest type Ia SNe data release.
It turns out that from all the possible scenarios,  the ones chosen by the best estimated values are \textit{precisely} the two scenarios $(\tilde{\zeta}_0 >0, \tilde{\zeta}_1<0)$ with $\tilde{\zeta}_0 + \tilde{\zeta}_1<3$, and $(0<\tilde{\zeta}_0 <3, \tilde{\zeta}_1=0)$.

The predicted values of the age of the Universe by the model when both $(\tilde{\zeta}_0, \tilde{\zeta}_1)$ are the free parameters, are a little smaller (11.72 and 10.03 Gyr) than the mean value of the observational constraint coming from the ages of the oldest galactic globular clusters but still inside of the confidence interval of this constraint ($12.9 \pm 2.9$ Gyr). However, when we set $\tilde{\zeta}_1=0$, the age of the Universe evaluate at the best estimate of $\tilde{\zeta}_0$ is 14.82~Gyr.

We find that the computed minimum values of the $\chi^2$-function by degrees of freedom (i.e., $\chi^2_{\rm
d.o.f.}$) are very close to the value one (or even smaller to one), indicating that the ``goodness-of-fit'' of the model to the SNe Ia data is excellent.

One of the disadvantages of the model parameterized by $\tilde{\zeta}=\tilde{\zeta}_0+\tilde{\zeta}_1H$,  is that the best estimated \textit{total} bulk viscosity function $\zeta(z)$  is positive for redshifts $z \lesssim 1$ but \textit{negative} for $z \gtrsim 1$. According to the local  entropy production, negative values of $\zeta$ imply a violation to the local second law of the thermodynamics. So, this is one of the drawbacks of the model.
We find that the cause of this violation is due the fact of having considered the existence of the term $\tilde{\zeta}_1H$ in the total bulk viscosity, this term characterizes a bulk viscosity proportional to the expansion ratio of the Universe.

For the case where we set  $\tilde{\zeta}_1=0$, we find that the total bulk viscosity is now \textit{positive} and  constant along the whole history of the Universe. This very simple model, in addition to the fact of that it does not violate the local entropy law, it has a good fit to the SNe Ia observations (measured through the $\chi^2_{\rm d.o.f.}$), the age of the Universe is in perfect agreement with the observational constraints and it predicts a Big-Bang with a decelerated expansion followed with a transition to an accelerated epoch. This simple model of one parameter $\tilde{\zeta}=\tilde{\zeta}_0$ is
a better candidate to explain the present accelerated expansion than the model with two parameters $\tilde{\zeta}=\tilde{\zeta}_0+\tilde{\zeta}_1H$.

Finally, it is important to mention that to have \textit{viable} bulk viscous  models, a physical mechanism  to explain the \textit{origin} of this bulk viscosity must be proposed. There are already some proposals in that sense, for instance, a decay of dark matter particles into relativistic product is claimed as explanation of its origin. A preliminary study of this propose, using the SNe Ia test, have been done in \cite{Wilson,Mathews2008} showing that
it could be a promising mechanism to explain this origin.

\end{document}